\documentclass[prd,aps,letterpaper,twocolumn,superscriptaddress,preprintnumbers,nofootinbib,floatfix]{revtex4-2}
\pdfoutput=1
\usepackage{xcolor}
\usepackage{graphicx}
\usepackage{amsmath}
\usepackage{amsfonts}
\usepackage{amssymb}
\usepackage{rotating}
\usepackage{subfigure}
\usepackage{paralist} 
\usepackage{verbatim}
\usepackage{float}
\usepackage{soul} 
\usepackage{appendix}
\usepackage{natbib}
\usepackage{epsfig}
\usepackage{hyperref}
\hypersetup{colorlinks,linkcolor=red,urlcolor=blue,citecolor=blue}
\usepackage[utf8]{inputenc}
\usepackage[english]{babel}
\usepackage{tabularx}
\newcolumntype{C}{>{\centering\arraybackslash}X}
\usepackage{mathpazo} % for math fonts
\usepackage{tgpagella} % overrides the text fonts in mathpazo
\usepackage{float}
\usepackage{listings} % for code block
\newcommand{\mnras}{Monthly Notices of the Royal Astronomical Society}
\newcommand{\apjl}{Astrophysical Journal Letters}
\newcommand{\apjs}{Astrophysical Journal Supplement}

\newcommand{\aap}{Astronomy \& Astrophysics}
\newcommand{\jcap}{J. Cosmol. Astropart. Phys.}

\usepackage{booktabs}
\usepackage{array, multirow}

\usepackage{lipsum}

\lstdefinestyle{mystyle}{
    basicstyle=\ttfamily\small,
    breakatwhitespace=true,         
    breaklines=true,   
}

\definecolor{cadetblue}{rgb}{0.37, 0.62, 0.63}

\newcommand{\stonybrook}{Physics and Astronomy Department, Stony Brook University, Stony Brook, NY  11794}

\begin{document}

\title{CMB-HD as a Probe of Dark Matter on Sub-Galactic Scales}

\author{Amanda MacInnis}
\affiliation{\stonybrook}

\author{Neelima Sehgal}
\affiliation{\stonybrook}

%%%%%%%%%%%%%%%%%%%%%%%%%%%%%%%%%%%%%%%%%%%%%%%%%%%%%%%
\begin{abstract}
We show for the first time that high-resolution CMB lensing observations can probe structure on sub-galactic scales.  In particular, a CMB-HD experiment can probe out to $k \sim 55~h$/Mpc, corresponding to halo masses of about $10^8 M_{\odot}$. Over the range $0.005~h$/Mpc$~< k <~55~h$/Mpc, spanning four orders of magnitude, the total lensing signal-to-noise ratio (SNR) from the temperature, polarization, and lensing power spectra is greater than 1900. CMB-HD gains most of the lensing SNR at small scales from the temperature power spectrum, as opposed to the lensing spectrum. These lensing measurements allow CMB-HD to distinguish between cold dark matter (CDM) and non-CDM models that change the matter power spectrum on sub-galactic scales. We also find that CMB-HD can distinguish between baryonic feedback effects and non-CDM models due to the different way each impacts the lensing signal. The kinetic Sunyaev-Zel'dovich (kSZ) power spectrum further constrains non-CDM models that deviate from CDM on the smallest scales CMB-HD measures. For example, CMB-HD can detect 1~keV warm dark matter (WDM) at  30$\sigma$, or rule out about 7~keV WDM at 95\%~CL, in a $\Lambda$WDM$+N_{\rm{eff}}+ \sum m_{\nu} + m_{\rm{WDM}} + {\rm{log_{10}}} T_{\rm{AGN}} + A_{\rm{kSZ}} + n_{\rm{kSZ}}$ model; here $T_{\rm{AGN}}$ characterizes the strength of the feedback, and $A_{\rm{kSZ}}$ and $n_{\rm{kSZ}}$ allow freedom in the amplitude and slope of the kinetic Sunyaev-Zel'dovich power spectrum. This work provides an initial exploration
of what can be achieved with reasonable assumptions about systematic effects. We make the CMB-HD Fisher code used here publicly available, and note that it can be modified to use any non-CDM model that changes the matter power spectrum.
\end{abstract}

\maketitle

\section{Introduction}
\label{sec:intro}
Mapping the matter distribution in the Universe is a key goal of modern cosmology.  Mapping the matter distribution on large scales probes the evolution of structure, which in turn yields knowledge of the Universe's components and history. Comparison of large-scale structure measurements with measurements from expansion-rate probes also informs us about the nature of gravity and dark energy. Mapping the matter distribution on small-scales, on the other hand, has become a critical probe of dark matter properties. Such measurements may be the only way to probe dark matter if dark matter does not interact with standard model particles.  

Measuring the gravitational lensing of the Cosmic Microwave Background (CMB) is a powerful way to measure the matter distribution~\cite{hu2001mappingDM,huokamoto2002,LewisChallinor2006}. One advantage of this method is that it probes dark matter directly via gravitational lensing, instead of relying on baryonic tracers (such as stars, galaxies, or gas).  Using the CMB as the background light source that undergoes lensing also has a number of unique advantages: 1)~the source light is from a well-known redshift, 2)~the unlensed properties of the source light are well understood, 3)~the source light is behind all structure in the Universe, and 4)~the source light on small scales is a smooth gradient field that retains the same gradient when viewed with infinite resolution. 

While the first direct detections of CMB lensing occurred relatively recently~\cite{Das2011,vanEngelen2012,Planck2013lensing}, CMB lensing measurements have rapidly increased in signal-to-noise ratio (SNR) on large scales~\cite{Planck2018lensing,ACTdr6lensing,Madhavacheril2023,Pan2023}, as well as on the modest scales of galaxy clusters~\cite{Baxter2014}, and galaxies~\cite{Madhavacheril2014}.  The large-scale measurements in particular are now routinely included when obtaining the tightest constraints on cosmological parameters~\cite[e.g.,][]{planck18params,DESI2024}.

The kinetic Sunyaev-Zel'dovich (kSZ) power spectrum is another measurement CMB observatories offer, which is sensitive to the distribution of dark matter on small scales~\cite{Farren2021}.  The kSZ effect arises when CMB photons scatter off electrons in ionized gas in the Universe and obtain a Doppler shift due to the bulk radial velocity of the gas~\cite{SZ1969,SZ1970,SZ1972}. Similar to CMB lensing, the kSZ spectrum can be measured both from the CMB temperature power spectrum and from the trispectrum~\cite{Smith2016}. Both the lensing and kSZ effects are frequency-independent, and can be separated from frequency-dependent foregrounds by CMB experiments that observe at many different frequencies.  While the kSZ power spectrum is routinely fit for in CMB power spectrum measurements~\cite{Planck2019Spectra,Choi2020,Balkenhol2022}, it has yet to be directly measured via the trispectrum; recent efforts to measure the kSZ trispectrum have yielded upper limits~\cite{Raghunathan2024,MacCrann2024}.  Measurements of the kSZ effect are expected to improve rapidly with future CMB experiments~\cite{Jain2023}.

In this work we quantify the ability of CMB-HD~\cite{HDastro2020,HDsnowmass} to measure the gravitational lensing of the CMB and the kSZ power spectrum. Here we provide a first look at what can be achieved with reasonable assumptions about systematic effects.  We stress that a more complete analysis of systematics is warranted and will be discussed in future work. We give special attention to the ability of CMB-HD to probe sub-galactic scales, which is a new regime for CMB observations.  The matter distribution on sub-galactic scales is of particular interest for understanding the nature of dark matter.  We also focus on the ability of CMB-HD measurements to distinguish between the effect of baryonic feedback (which can move around the distribution of matter on small-scales), and the effect of dark matter that differs from cold dark matter (CDM).  Distinguishing between these two scenarios is necessary to probe the behavior of dark matter alone. Lastly, we quantify the extent to which varying the standard $\Lambda$CDM parameters in addition to extensions (e.g.~$N_{\rm{eff}}$ and $\sum m_{\nu}$) impacts constraints on dark matter properties. Given the enormous precision with which CMB-HD can measure the primordial CMB, CMB lensing over four orders of magnitude in scale, and the kSZ power spectrum, we examine whether CMB-HD could adopt a holistic approach of varying all relevant parameters simultaneously.

In Section~\ref{sec:key-result}, we briefly summarize the key results of this work.  In Section~\ref{sec:methods} we discuss how we calculate covariance matrices, CMB and lensing spectra, lensing SNRs, the effect of non-CDM models, the kSZ spectrum, and the Fisher matrix.  In Section~\ref{sec:results}, we present our results, and in Section~\ref{sec:discussion}, we discuss and conclude.

\vspace{-4mm}
\section{Summary of Key Results}
\label{sec:key-result}

\vspace{-4mm}
Below we briefly summarize the key results. 

\vspace{-1mm}
\begin{itemize}
    \item We forecast the SNR for measuring lensing from the CMB-HD $TT$, $TE$, $EE$, $BB$, and $\kappa\kappa$ power spectra and show the results in Table~\ref{tab:lensing_snr}.  We find that the total lensing SNR for all the spectra combined is 1947, and that more SNR comes from the temperature, as opposed to the lensing, power spectra.  

    \vspace{-1mm}
    \item The top panel of Figure~\ref{fig:SNR} shows the breakdown of the lensing SNR by wavenumber $k$ (in~$h$/Mpc) and by redshift. It shows that smaller-scale signals originate from lower redshifts. It also shows that CMB-HD lensing measurements would span $0.005~h$/Mpc$ < k < 55~h$/Mpc, over four orders of magnitude in scale, with a SNR in the several hundreds for most of the $k$ bins shown. The bottom panel of Figure~\ref{fig:SNR} further divides into $\ell$ bins, and indicates that CMB-HD would probe lensing well into the non-linear regime.

    \vspace{-1mm}
    \item In Figure~\ref{fig:Pk}, we show forecasted CMB-HD constraints on the {\it{linear}} matter power spectrum following~\cite{Chabanier2019} and~\cite{Tegmark2002}. We find that CMB-HD can measure the matter power spectrum out to $k\sim~55~h$/Mpc, corresponding to halos of $10^8~M_{\odot}$.  

    \vspace{-1mm}
    \item We show in Table~\ref{tab:WDMerror} the forecasted constraints on the warm dark matter (WDM) mass ($m_{\rm{WDM}}$) using CMB-HD power spectra plus DESI baryon acoustic oscillation (BAO) data~\cite{desi}.  We find that $m_{\rm{WDM}}$ is most degenerate with the slope and amplitude of the kSZ power spectrum, $n_{\rm{kSZ}}$ and $A_{rm{kSZ}}$, and that $m_{\rm{WM}}$ constraints are significantly degraded when these parameters are allowed to be free.  This degeneracy is shown in Figures~\ref{fig:Triangle} and~\ref{fig:TriangleSmall}.

    \vspace{-1mm}   
    \item  We find that varying the $\Lambda$CDM parameters, the number of relativistic species ($N_{\rm{eff}}$), and the sum of the neutrino masses ($\sum m_{\nu}$) negligibly impacts $m_{\rm{WDM}}$ constraints, as shown in Table~\ref{tab:WDMerror}. From Figures~\ref{fig:Triangle} and~\ref{fig:TriangleSmall}, we see that the most notable, but still mild, degeneracies are with $n_\mathrm{s}$ and $N_{\rm{eff}}$.

    \vspace{-1mm}
    \item As shown in Figures~\ref{fig:Triangle} and~\ref{fig:TriangleSmall} and Table~\ref{tab:WDMerror}, marginalizing over baryonic feedback effects (${\rm{log_{10}}} T_{\rm{AGN}}$) does not degrade constraints on $m_{\rm{WDM}}$; this is due to the substantially different way baryonic effects and warm dark matter alter the matter power spectrum, shown via the difference in the CMB lensing signal in Figure~\ref{fig:TT_kk_ratio}. 

\begin{figure}[t]
    \centering
    \includegraphics[width=\columnwidth]{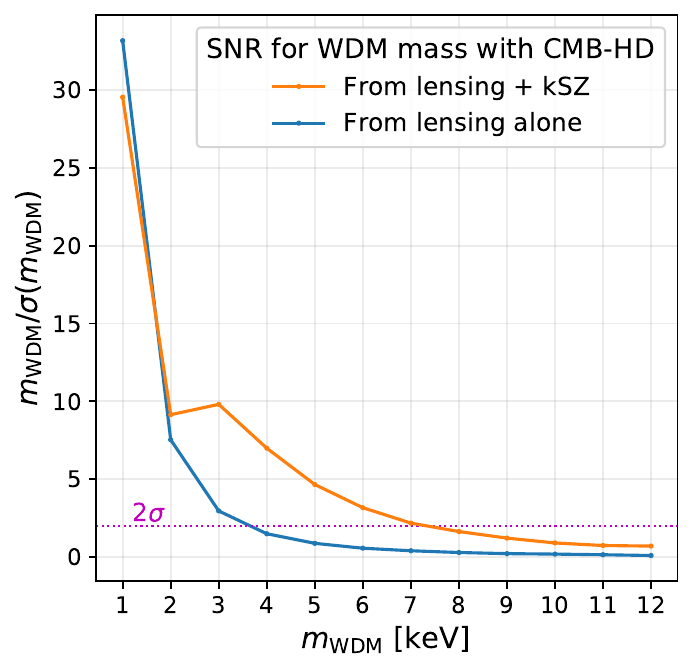}
    \caption{We show the significance with which CMB-HD plus DESI BAO can measure a given warm dark matter (WDM) mass, quantified by the signal-to-noise ratio (SNR) $m_\mathrm{WDM} / \sigma(m_\mathrm{WDM})$. The blue curve shows the SNR obtained from CMB-HD lensing data alone, while the orange curve uses both CMB lensing and kSZ data. We assume a $\Lambda$WDM$+N_{\rm{eff}}+ \sum m_{\nu} + m_{\rm{WDM}} + {\rm{log_{10}}} T_{\rm{AGN}}$ model when only using lensing data, and add kSZ parameters $A_{\rm{kSZ}}$ and $n_{\rm{kSZ}}$ to the model when adding kSZ data; all parameters are varied simultaneously.  We use the Warm\&Fuzzy code of~\protect{\cite{Marsh2016}} to calculate the non-linear WDM transfer function (Eq.~\ref{eq:transfer}). Given the significant challenge of modeling this transfer function even for WDM models, the plot above is illustrative of what CMB-HD can measure; however, exact SNRs will depend on modeling details.  We find that CMB-HD can detect WDM with a mass of about 7~keV at the 95\% C.L. when including kSZ measurements, or 3.5~keV WDM from lensing alone.} 
    \vspace{-3mm} 
    \label{fig:wdm_mass_snr}
\end{figure}
    
    \vspace{-1mm} 
    \item We find that CMB-HD plus DESI BAO can detect 1~keV WDM at 30$\sigma$, or rule out about 7~keV WDM at 95\%~C.L., in a $\Lambda$WDM$+N_{\rm{eff}}+ \sum m_{\nu} + {\rm{log_{10}}} T_{\rm{AGN}} + A_{\rm{kSZ}} + {n_{\rm{kSZ}}} +  m_{\rm{WDM}}$ model, which we illustrate in Figure~\ref{fig:wdm_mass_snr}. 

    \vspace{-1mm}
    \item We make public the code to reproduce the results in this work\footnote{\url{https://github.com/CMB-HD/hdPk}}. We note that it can be modified to use any non-CDM model that changes the non-linear matter power spectrum. 
    
\end{itemize}

\vspace{-2mm}
\begin{figure*}[t]
    \centering
    \includegraphics[width=0.95\textwidth]{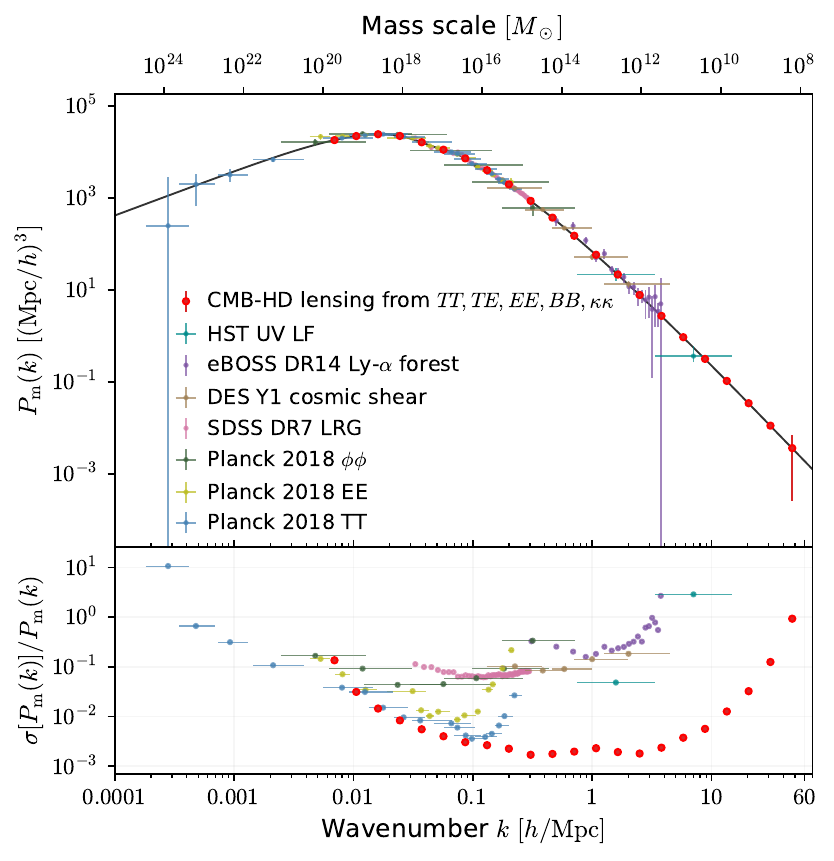} 
    \caption{\textit{Top panel}: We illustrate how CMB-HD lensing measurements can probe the \textit{linear} matter power spectrum. We show in red the forecasted errors; these are obtained from the CMB-HD lensing SNR in each comoving wavenumber bin,  equating the inverse of this lensing SNR to the fractional error bar on the linear matter power spectrum (see Section~\ref{sec:snr-perk} for SNR calculation and Eq.~\ref{eq:PkError} in Section~\ref{sec:lensing-snr}). We place the error bars on the theoretical prediction for the linear matter power spectrum (black curve) at the center of each wavenumber bin. We also indicate on the top $x$-axis the corresponding mass scale today for the comoving wavenumbers on the bottom $x$-axis, using Eq.~\ref{eq:mass}. Of note is that the CMB-HD lensing constraints span four orders of magnitude in wavenumber, $0.005~h$/Mpc$~< k <~55~h$/Mpc, and 12 orders of magnitude in mass scale, going down to $10^8 M_{\odot}$. With the exception of the last bin, the error bars are smaller than the band center markers. We also include for reference constraints already measured and derived by~\protect{\cite{Chabanier2019}} based on the method of~\protect{\cite{Tegmark2002}}: Lyman-$\alpha$ data~\protect{\cite{Chabanier2018}} from the eBOSS DR14 release~\protect{\cite{eBOSSdr14}} (purple); cosmic shear data from DES Y1~\protect{\cite{DESY1shear}} (brown); galaxy clustering data for luminous red galaxies (LRG) from the SDSS DR7 release~\protect{\cite{sdssDR7LRG}} (pink); and \textit{Planck} 2018 CMB temperature (blue), polarization (yellow), and lensing (green) data~\protect{\cite{Planck2018overview,Planck2018lensing}}. We also add constraints derived from high-redshift measurements of the UV galaxy luminosity function (HST UV LF) from~\protect{\cite{Sabti2021}} (teal). Note that this is not an exhaustive summary of current constraints, but illustrative of what has already been achieved.  We include data that more directly measures the matter power spectrum, as opposed to data that measures the halo mass function that is then used to infer the linear matter power spectrum, which requires first marginalizing over cosmology, baryonic physics, and sample variance, and modeling very nonlinear scales.
    \textit{Bottom panel}: We compare the error bars on the matter power spectrum from each data set as a fraction of the total power. We see that CMB-HD achieves sub-percent fractional errors in the range $0.03 ~h/$Mpc~$<~k~<~10$~$h/$Mpc. }
    \label{fig:Pk}
\end{figure*}

\section{Methods}
\label{sec:methods}

We describe below the methods to calculate covariance matrices (Sections~\ref{sec:covmat} and~\ref{sec:desi}), signal spectra (Section~\ref{sec:signal}), lensing SNRs (Section~\ref{sec:snr}), the effects of non-CDM models (Section~\ref{sec:wdm-baryons}), the kSZ spectrum (Section~\ref{sec:kSZ}), and Fisher matrices (Section~\ref{sec:fisher}).

\subsection{CMB-HD Covariance Matrix}
\label{sec:covmat}

To calculate the CMB-HD covariance matrix we follow the procedure in~\cite{HDparams}. In Table~\ref{tab:ExpConfig}, we list the experimental configuration assumed.   To summarize, we include CMB and CMB lensing power spectra in the multipole range of $\ell \in [30, 20000]$ for $TT$, $TE$, $EE$, $BB$, and $\kappa\kappa$, and calculate the $5 n_\mathrm{bin} \times 5 n_\mathrm{bin}$ covariance matrix, $\mathbb{C}$, as in~\cite{HDparams}.  In addition, we extend the $\ell_\mathrm{max}$ for $TT$ to 40,000, which differs from~\cite{HDparams}. We calculate the covariance matrix, SNR, and Fisher matrix for $TT$ in the range of $\ell \in [20000, 40000]$ separately, as described below.

\subsubsection{Astrophysical Foregrounds}
\label{sec:fg}

\begin{figure}[t]
    \centering
    \includegraphics[width=\columnwidth]{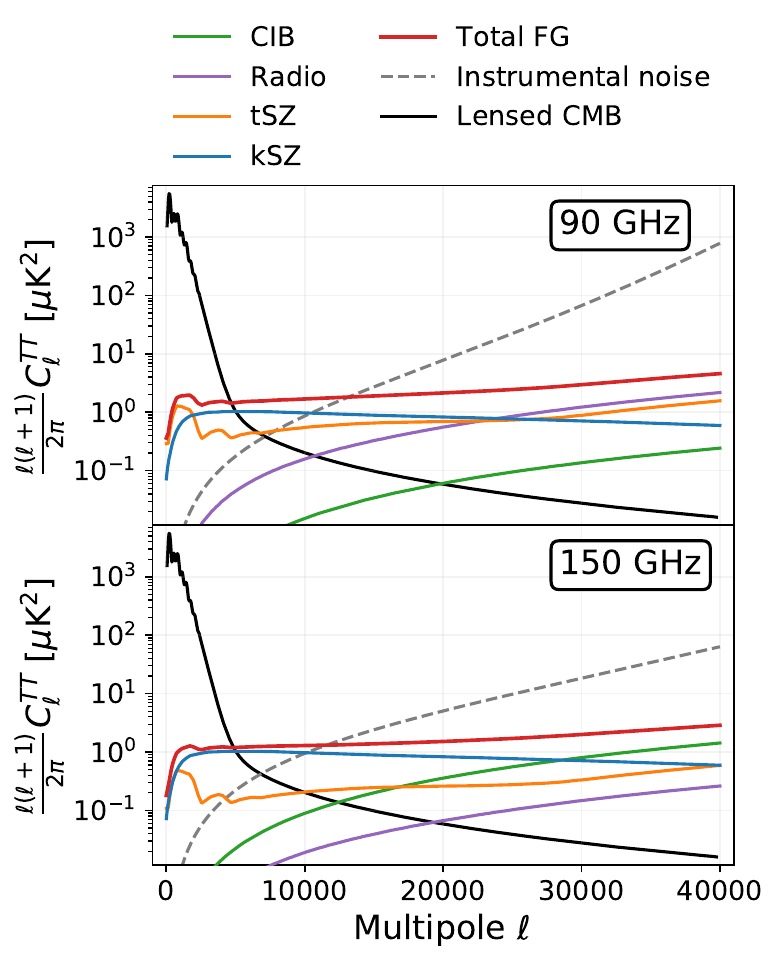}
    \caption{We show the residual extragalactic foreground power expected for CMB-HD at 90~GHz (top panel) and 150~GHz (bottom panel) after foreground cleaning (described in Section~\ref{sec:fg} and Appendix~\ref{sec:extragalacticFGs}), as well as the instrumental noise (gray dashed) and the lensed CMB temperature power spectrum (black). The residual power spectra of infrared (CIB) and radio sources are shown in green and purple, respectively, assuming that point sources detected at $5\sigma$ significance can be subtracted from the maps. In particular, we assume that radio sources with fluxes above 0.04~mJy can be detected at $5\sigma$ in the 90~GHz maps, corresponding to a flux limit of 0.03~mJy at 150~GHz. We assume that CIB sources with fluxes above 0.15~mJy can be detected at $5\sigma$ in the 280 GHz maps, corresponding to flux limits of 0.008~mJy and 0.03~mJy at 90~and 150~GHz, respectively. The residual CIB power also accounts for some source mis-subtraction due to uncertainty in the CIB spectral index. The residual thermal Sunyaev-Zel’dovich (tSZ) power (orange) is found by assuming that tSZ clusters detected above $3\sigma$ are subtracted from the maps. The CIB, radio, and tSZ power spectra were derived by~\protect{\cite{han22}}. The kinetic Sunyaev-Zel’dovich (kSZ) power, shown in blue, includes components from the epoch of reionization and from late times, calculated using templates from~\protect{\cite{Smith2018,Park2013}} and~\protect{\cite{Battaglia2010}}, respectively; we do not assume that this signal will be removed. The total extragalactic foreground power is in red. } 
    \label{fig:HDFGs}
\end{figure}

In this work, we account for the residual extragalactic foreground signal that will remain in the CMB-HD maps after foreground-cleaning procedures. Our foreground model includes contributions from the thermal and kinetic Sunyaev-Zel’dovich effects (tSZ and kSZ, respectively), the cosmic infrared background (CIB), and radio sources. The residual tSZ, CIB, and radio power spectra are obtained from~\cite{han22}, and we include the full power spectrum of the kinetic Sunyaev-Zel’dovich (kSZ) effect (with both reionization and late-time contributions) using a template from~\cite{Battaglia2010} for the late-time part, and from~\cite{Smith2018,Park2013} for the reionization part, normalized as described in Section~\ref{sec:kSZ}.\footnote{The model for the kSZ power spectrum used in this work differs from that in~\cite{han22} and~\cite{HDparams}, by including both the late-time and reionization kSZ contributions, as opposed to just the latter. However, the contribution of small-scale $TT$ data to our lensing noise (Section~\ref{sec:LensingNoise}) and covariance matrix (Section~\ref{sec:CovmatConstruction}) is derived from the covariance matrix of~\cite{han22}, which only includes the reionization kSZ component.} We show these residual foreground levels in Figure~\ref{fig:HDFGs} for the 90~and 150~GHz channels, which are the frequencies that contain the most statistical weight in cosmological parameter analysis. We describe our extragalactic foreground model in detail in Appendix~\ref{sec:extragalacticFGs}.

\begin{table}[t]
    \begin{center}
    \begin{tabular}{cccccr}
      \toprule
      \toprule
      \multirow{2}{*}{Exp.} & \multirow{2}{*}{$f_\mathrm{sky}$} &  Freq., & Temp. Noise, & FWHM,  &  \multicolumn{1}{c}{Multipole Range}
      \\
      & &   GHz & $\mu$K-arcmin & arcmin  & \multicolumn{1}{c}{$\ell, L$}
      \\
      \midrule
      \multirow{3}{*}{HD} & \multirow{3}{*}{0.6} &  90 & 0.7 & 0.42 & TT: [30,\ 40000]
      \\
      &  &  150  & 0.8  & 0.25  & TE, EE, $\kappa\kappa$: [30,\ 20000]
      \\
      &  &   &   &   & BB:  [1000,\ 20000]
      \\
      \midrule[0.2pt]
      \multirow{2}{*}{ASO} & \multirow{2}{*}{0.6} & 90 & 3.5 & 2.2 & \multirow{2}{*}{BB: [30,1000)}
      \\
      & & 150 & 3.8 & 1.4 & 
      \\
      \bottomrule
    \end{tabular}
    \caption{We list white noise levels and beam full-width at half-maximum (FWHM) at 90 and 150 GHz for CMB-HD (``HD'') and the Advanced Simons Observatory (``ASO''), along with the assumed sky fraction ($f_\mathrm{sky}$) that each experiment will observe, and the multipole ranges of the power spectra from each experiment we use in our forecasts. The map noise levels given are for temperature; the noise levels for polarization are  higher by a factor of $\sqrt{2}$. The CMB-HD experimental specifications are from~\protect{\cite{HDsnowmass}}.}
    \label{tab:ExpConfig}
    \end{center}
\end{table}

As discussed below, we assume that large-scale CMB modes below $\ell=1000$, will be measured by the Advanced Simons Observatory (ASO) and not by CMB-HD.  {\it{Galactic}} foregrounds will be an issue on these scales and we assume the many frequency channels of ASO will be enough to clean these foregrounds to below the ASO instrumental noise levels.  Since ASO is targeting an ambitious science goal of measuring primordial gravitational waves, subtracting Galactic foregrounds is a high priority, and thus we expect that it will be done sufficiently and do not include it in our noise model for large-scales.

Regarding atmospheric noise fluctuations, they mainly impact large scales ($\ell<1000$).  For ASO, the atmospheric noise is expected to be below the signal in 90 and 150 GHz channels down to $\ell=30$ for $TT, TE$, and $EE$~\cite{SOforecasts}.   Recent work modeling atmospheric fluctuations for large-single dish experiments ($\ge 30$ meters), such as CMB-HD, suggests that these experiments have an advantage in removing atmospheric fluctuations due to the enhanced correlation of atmospheric effects across beams of adjacent detectors~\cite{vanMarrewijk2024}. Thus, we do not include atmospheric noise in our calculations below.  We leave for future work modeling atmospheric effects given CMB-HD's specific observing strategy, and including it in a more refined noise model.

There are also instrumental systematics that impact CMB experiments. Some of these include beam systematics, gain calibration, polarization efficiencies etc. Including these details is beyond the scope of this initial investigation and will depend on the exact instrumental design.  However, eventually they should be included in future work to achieve more precise forecasts.

\subsubsection{Noise on CMB Power Spectra}
\label{sec:CMBnoise}

We assume CMB-HD will observe 60\% of the sky, and that the 90 and 150 GHz channels will be the main contributors to the parameter analysis (while the other five frequency channels will mainly be used to reduce foreground contamination). We list the instrument white noise levels and beam full width at half-maximum (FWHM) at 90 and 150 GHz for CMB-HD in Table~\ref{tab:ExpConfig}~\cite{HDsnowmass}. We assume that the temperature and polarization noise are uncorrelated.

We assume that CMB-HD will measure multipoles of $\ell \geq 1000$, and that CMB data in the range $30 \leq \ell < 1000$ will come from measurements by precursor CMB surveys. Since $TT$, $TE$, and $EE$ will already be sample-variance limited over 60\% of the sky for $30 \leq \ell < 1000$ from precursor surveys such as the Simons Observatory (SO)~\cite{SOforecasts}, we extend the CMB-HD multipole range down to $\ell_\mathrm{min} = 30$ for these spectra since CMB-HD will have full overlap with these surveys. For the $BB$ power spectrum, we use the expected noise from the ASO in the range $\ell \in [30, 1000)$. We provide the expected instrumental noise levels in temperature and the beam FWHM at 90 and 150 GHz for ASO in the last row of Table~\ref{tab:ExpConfig}.

We also include the effect of residual extragalactic foregrounds in the temperature data by adding the expected residual foreground power spectra for CMB-HD, described in Section~\ref{sec:fg} and Appendix~\ref{sec:extragalacticFGs}, to the temperature instrument noise spectra at 90 and 150 GHz. The coadded noise spectra, $N_\ell^{XY}$, for $XY \in \{TT, TE, EE, BB\}$ are obtained by calculating the beam-deconvolved instrumental noise power spectrum for each frequency channel (90 and 150 GHz), adding the residual foreground power to the beam-deconvolved instrumental noise for $TT$ for each frequency, and coadding the spectra of the two frequencies using inverse-noise weighting.

\subsubsection{Noise on CMB Lensing Spectra}
\label{sec:LensingNoise}

The gravitational lensing of the primary CMB introduces coupling between modes of the lensed CMB on different scales. This lensing-induced mode coupling can be reconstructed from pairs of CMB maps that are filtered to isolate the non-Gaussian effects of lensing, forming a quadratic estimator~\cite{hu2001mappingDM,huokamoto2002,OkamotoHu2003}. The power spectrum of this estimator depends on the four-point function of the CMB maps used to construct it, which has two components~\cite{Hu2001trispectrum}. The non-Gaussian part of the four-point function contains the CMB lensing power spectrum, but the Gaussian component introduces a bias to the lensing power spectrum, which arises from the primordial CMB signal and the instrumental noise. This Gaussian bias, referred to as the $N^{(0)}$ bias, can be removed using a realization-dependent (RDN0) subtraction technique~\cite{Namikawa2013}; however, it will still act as a source of noise on the estimated CMB lensing power spectrum.

We use the lensing noise $N_\ell^{\kappa\kappa}$ from~\cite{HDparams}, which was calculated using two kinds of quadratic estimators. We briefly discuss the calculation here, and refer the reader to~\cite{HDparams} for more details. 

The first estimator is the ``standard'' quadratic estimator from~\cite{hu2001mappingDM,huokamoto2002,OkamotoHu2003}, which we refer to as the H\&O estimator. It is designed to minimize the variance of the lensing reconstruction. We use the CLASS delens package~\cite{Hotinli2021}\footnote{\url{https://github.com/selimhotinli/class_delens}} to calculate $N_\ell^{\kappa\kappa}$ for the $TT$, $TE$, $EE$, $EB$, and $TB$ H\&O estimators, which uses an iterative delensing technique to reduce their noise levels. This calculation includes the residual extragalactic foreground power for temperature spectra mentioned above, and the ASO instrumental noise in polarization on scales $\ell < 1000$.  

The second estimator is the quadratic estimator of~\cite{hdv2007}, which we refer to as the HDV estimator.  It is designed to more optimally reconstruct small scales. Here, as in~\cite{HDparams} and~\cite{han22}, we also use it to reduce bias to the lensing reconstruction on small-scales that arises due the presence of non-Gaussian foregrounds. This is done by filtering one CMB map to isolate large scales ($\ell < 2000$), and the other CMB map to isolate only small scales ($\ell \in [5000, 20000]$) (see~\cite{han22} for details). We replace the noise from the H\&O $TT$ estimator on scales $\ell > 5000$ with the noise from the HDV $TT$ estimator, derived from the simulation-based covariance matrix of~\cite{han22}. These simulations include extragalactic foregrounds that are correlated with the non-Gaussian CMB lensing map~\cite{han22}, using the  anticipated residual foreground levels for CMB-HD described in Section~\ref{sec:fg} and Appendix~\ref{sec:extragalacticFGs} (with the exception of the kSZ effect, which only includes the reionization component as mentioned above).

The final lensing noise power spectrum $N_\ell^{\kappa\kappa}$ is a minimum-variance combination of the lensing power spectrum noise curves from the $TT$, $TE$, $EE$, $EB$, and $TB$ lensing estimators.

\subsubsection{Construction of lensed and delensed covariance matrices} 
\label{sec:CovmatConstruction}

We follow a similar approach as~\cite{HDparams} to calculate the covariance matrices for the lensed and delensed CMB spectra and CMB lensing spectrum in the multipole range $\ell \in [30, 20000]$. Each of the lensed and delensed covariance matrices is a $5 n_\mathrm{bin} \times 5 n_\mathrm{bin}$ matrix, since five spectra are used in the forecasts. The full lensed or delensed covariance matrix includes diagonal terms to account for the Gaussian variance of the spectra, and off-diagonal terms to capture the non-Gaussian lensing-induced covariances between the spectra. We summarize the calculation in Appendix~\ref{sec:HDcovmatcalc}, which is discussed in more detail in~\cite{HDparams}, \cite{Hotinli2021}, \cite{BenoitLevySmithHu2012}, and~\cite{han22}.

\subsection{DESI Covariance Matrix} \label{sec:desi}

To forecast parameters from the combination of CMB-HD and DESI BAO, we also estimate the DESI covariance matrix.  We simulate a DESI dataset of BAO measurements taken over 14,000 square degrees of the sky from the baseline galaxy survey ($z \in [0.65, 1.85]$) and bright galaxy survey ($z \in [0.05, 0.45]$)~\cite{desi}.  The BAO data consists of the distance ratio measurements $r_\mathrm{s} / d_V(z)$ at redshift $z$, where $r_\mathrm{s}$ is the comoving sound horizon at the end of the baryon drag epoch and the combined distance measurement $d_V(z)$ is defined as~\cite{Eisenstein2005}
\begin{equation} \label{eq:BAOdV}
    d_V(z) \equiv \left[D_M(z)^2 \frac{cz}{H(z)}\right]^{1/3}.
\end{equation}
Here $D_M(z) = (1+z) d_A(z)$, $d_A(z)$ is the angular diameter distance to redshift $z$, and $H(z)$ is the expansion rate at redshift $z$. We use CAMB to calculate the theoretical $r_\mathrm{s} / d_V(z_i)$ at each redshift $z_i$. We follow the same approach as in~\cite{HDparams} to form the covariance matrix for the DESI BAO data, by assuming it is diagonal and propagating the fractional errors on $r_\mathrm{s}/d_A(z_i)$ and $r_\mathrm{s} H(z_i)$ forecasted by~\cite{desi} into an uncertainty $\sigma_i$ on $r_\mathrm{s} / d_V(z_i)$ at each redshift $z_i$. The diagonal elements of the covariance matrix are then taken to be $\sigma_i^2$.

\subsection{Calculating CMB and CMB Lensing Signal Spectra}
\label{sec:signal}

While CAMB~\cite{CAMBHowlett:2012mh, CAMBLewis:1999bs} can calculate the lensed CMB and the CMB lensing power spectra in a $\Lambda$CDM model (with the option to include the effects of baryonic feedback using the HMcode2020 model~\cite{Mead2020}), it does not calculate these spectra for an arbitrary non-CDM model that changes the non-linear matter power spectrum.  Thus, we calculate the CMB lensing power spectrum outside of CAMB as described below, and then input this into CAMB to get the lensed and delensed CMB power spectra.  We show in Appendix~\ref{sec:CAMBComparison} that for a $\Lambda$CDM model, our calculation agrees with the CAMB-only calculation to within 0.3\% for $\ell \leq 20,000$, and to within 0.8\% for $\ell \leq 40,000$.

To calculate the lensing power spectrum we follow~\cite{LewisChallinor2006} and use the Limber approximation to obtain
\begin{equation} \label{eq:ClPhiPhi}
    C_\ell^{\phi\phi} = 4 \int_0^{\chi_s} d\chi \left(\frac{\chi_s - \chi}{ \chi^2\chi_s}\right)^2 P_\Psi\left(k=\frac{\ell + 1/2}{\chi}, z(\chi)\right).
\end{equation}
Here $k \approx (\ell + 1/2) / \chi(z)$, where $k$ is the comoving wavenumber in Mpc$^{-1}$,\footnote{Note that we plot the comoving wavenumber $k$ in units of $h$ Mpc$^{-1}$ in the figures throughout this work as is standard.} and $\chi(z)$ is the comoving radial distance to redshift $z$  in Mpc~\cite{LewisChallinor2006,Dodelson2020}. $\chi_s = \chi(z_s)$ is the comoving distance to the last scattering surface at redshift $z_s \approx 1100$. The CMB lensing convergence power spectrum $C_\ell^{\kappa\kappa}$ is related to this by 
\begin{equation} \label{eq:phi2kappa}
    C_\ell^{\kappa\kappa} = \frac{[\ell(\ell+1)]^2}{4} C_\ell^{\phi\phi}.
\end{equation}

The power spectrum $P_\Psi(k,z)$ of the three-dimensional gravitational potential $\Psi(\mathbf{k},z)$ is related to the non-linear matter power spectrum $P_\mathrm{m}(k,z)$ by~\cite{Dodelson2020,Nguyen2017,LewisChallinor2006}
\begin{equation} \label{eq:PsiToMatterPower}
    P_\Psi(k,z) = \left(\frac{3 \Omega_\mathrm{m} H_0^2}{2 c^2}\right)^2 \frac{(1+z)^2}{k^4} P_\mathrm{m}(k,z),
\end{equation}
where $\Omega_\mathrm{m}$ and $H_0$ are the matter density and Hubble rate today. Thus, $C_\ell^{\phi\phi}$ can also be expressed as an integral over $P_\mathrm{m}(k,z)$.  

To calculate the CMB lensing power spectrum for a non-CDM model, we apply a transfer function to the non-linear matter power spectrum, defined as 
\begin{equation} \label{eq:transfer}
    T^2_\mathrm{non-lin}(k,z) = \frac{P_\mathrm{m}^\mathrm{Non-CDM}(k,z)}{P_\mathrm{m}^\mathrm{CDM}(k,z)}.
\end{equation}
Given the relationship between $P_\Psi$ and $P_\mathrm{m}$ in Eq.~\ref{eq:PsiToMatterPower}, this is equivalent to changing $P_\Psi(k,z)$ to $T^2(k,z) P_\Psi^\mathrm{CDM}(k,z)$ in the integral of Eq.~\ref{eq:ClPhiPhi}.  We use CAMB to calculate $P_\Psi^\mathrm{CDM}(k,z)$ with the accuracy settings described in~\cite{HDparams}.\footnote{We use the 2016 version of HMCode~\cite{Mead2016} to calculate the non-linear $P_\Psi^\mathrm{CDM}(k,z)$, and use the single-parameter baryonic feedback model of HMCode2020~\cite{Mead2020} to calculate $P_\Psi^\mathrm{CDM+baryons}(k,z)$ when including baryonic feedback.} For consistency we calculate $C_\ell^{\phi\phi}$ from Eq.~\ref{eq:ClPhiPhi} for all models, including $\Lambda$CDM and $\Lambda$CDM plus baryonic feedback.  

After we obtain $C_\ell^{\kappa\kappa}$, we provide it to CAMB so that CAMB can calculate the lensed $TT$, $TE$, $EE$, and $BB$ CMB power spectra from the corresponding unlensed spectra. The delensed power spectra are calculated in the same way as the lensed power spectra, except we replace $C_\ell^{\kappa\kappa}$ with the residual lensing power spectrum $C_\ell^{\kappa\kappa,\mathrm{res}}$, given by~\cite{han20delensing,HDparams} 
\begin{equation} \label{eq:ResidualLensingPower}
    C_\ell^{\kappa\kappa,\mathrm{res}} = C_\ell^{\kappa\kappa} \left(1 - \frac{C_\ell^{\kappa\kappa}}{C_\ell^{\kappa\kappa} + N_\ell^{\kappa\kappa}}\right).
\end{equation}
Here, $N_\ell^{\kappa\kappa}$ is the expected noise on the lensing power spectrum discussed above.

\subsection{Calculating Lensing Signal-to-Noise Ratios}
\label{sec:snr}

In order to quantify how well CMB-HD will be able to measure the gravitational lensing of the CMB, we use the SNR statistic defined as

\begin{equation} \label{eq:SNR}
        \frac{S}{N} = \sqrt{\sum_{ij} \left(\Delta C_{\ell}\right)_i \mathbb{C}^{-1}_{ij} \left(\Delta C_{\ell}\right)_j}.
\end{equation}
Here $\Delta C_{\ell}$ is a one-dimensional vector holding the difference between the two sets of model power spectra being considered, in this case lensed and unlensed spectra (with the unlensed $BB$ and $\kappa\kappa$ spectra set to zero).  Each spectrum is binned into $n_\mathrm{bin}$ bins, and the elements of $\Delta C_{\ell}$ are $\Delta C_{\ell} = (\Delta C_{\ell_1}^{TT}, \dots, \Delta C_{\ell_{n_\mathrm{bin}}}^{TT}, \Delta C_{\ell_1}^{TE}, \dots, \Delta C_{\ell_{n_\mathrm{bin}}}^{\kappa\kappa})$, in the order $TT$, $TE$, $EE$, $BB$, $\kappa\kappa$, where $\ell_b$ is the center of the $b^\mathrm{th}$ bin.  The sum in Eq.~\ref{eq:SNR} is taken over bin indices $i$ and $j$ from $1$ to $5 n_\mathrm{bin}$. $\mathbb{C}$ is the $5 n_\mathrm{bin} \times 5 n_\mathrm{bin}$ covariance matrix for the lensed spectra, with $n_\mathrm{bin} \times n_\mathrm{bin}$ blocks of $\mathbb{C}_{\ell_b \ell_{b'}}^{XY,WZ}$, where $XY$ and $WZ$ can each be one of $TT$, $TE$, $EE$, $BB$, or $\kappa\kappa$. The blocks are arranged in the same order as the spectra. This SNR statistic can also be applied to an individual spectrum $XY \in \{TT,TE,EE,BB,\kappa\kappa\}$ using the difference vector $\Delta C_\ell^{XY}$ of length $n_\mathrm{bin}$ and the $n_\mathrm{bin} \times n_\mathrm{bin}$ covariance matrix block $\mathbb{C}^{XY,XY}$.

\subsubsection{Lensing Signal-to-Noise Ratio per $k$ Mode}
\label{sec:snr-perk}

We follow the procedure discussed above, using Eq.~\ref{eq:SNR} and the covariance matrix described in Section~\ref{sec:covmat}, to calculate the SNRs given in Table~\ref{tab:lensing_snr} in Section~\ref{sec:results}.

However, {\it{for visualization purposes}}, it is useful to see what comoving wavenumber $k$ in units of $h$ Mpc$^{-1}$ the CMB-HD lensing signal is sensitive to (as we show in Figures~\ref{fig:Pk} and~\ref{fig:SNR}).  Below, we summarize how we approximate this, giving full details in Appendix~\ref{sec:SNRapprox}.  

The lensing signal in the CMB maps provides a measurement of the matter distribution, integrated along the line of sight over all redshifts from recombination until today. As discussed in Section~\ref{sec:signal}, we use the relationship $k \approx (\ell + 1/2) / \chi(z)$ to translate the lensing measured from the CMB power spectra on a given angular scale (multipole $\ell$) to a measurement of the matter distribution on a given physical scale (wavenumber $k$) at a given redshift $z$. Therefore, to approximate the lensing SNR from a given wavenumber bin, we first need to calculate the SNR per multipole in each redshift bin.   

We calculate the contribution to the lensing signal from a given redshift bin of width $\Delta z = 0.5$, for redshifts $z \in [0,6]$.\footnote{While the lensing signal is from all redshifts $z \in$ [0, 1100], we only go to $z=6$ in Figures~\ref{fig:Pk} and \ref{fig:SNR} to keep our redshift bins small and ease the conversion from $\ell$ to $k$ at a given redshift.  We note that the range $z \in [6, 1100]$ contributes only 15\% extra lensing signal.} The $i^\mathrm{th}$ redshift bin ranges from $z_{i,\mathrm{min}}$ to $z_{i,\mathrm{max}}$, centered at $z_i$.  We integrate Eq.~\ref{eq:ClPhiPhi} from $\chi(z_{i,\mathrm{min}})$ to $\chi(z_{i,\mathrm{max}})$ and then use Eq.~\ref{eq:phi2kappa} to calculate $C_\ell^{\kappa\kappa, z_i}$, which contains the lensing signal from only the $i^\mathrm{th}$ redshift bin. We use CAMB to obtain the lensed CMB spectra $C_\ell^{UV, z_i}$, for $UV \in \{TT,TE,EE,BB\}$, that were lensed by this $C_\ell^{\kappa\kappa,z_i}$. We then use the set of $C_\ell^{XY,z_i}$, for $XY \in \{TT,TE,EE,BB,\kappa\kappa\}$, in Eqs.~\ref{eq:GaussCovCMB} and~\ref{eq:GaussCovLensing} to calculate the diagonal elements of the covariance matrices for the five spectra, $\left(\sigma_{\ell}^{XY,z_i}\right)^2$. In Eqs.~\ref{eq:GaussCovCMB} and~\ref{eq:GaussCovLensing}, $N_\ell^{XY}$ is the noise on the power spectrum, which we assume is the same for each redshift bin. The lensing SNR per multipole $\ell$ from the $i^\mathrm{th}$ redshift bin is then defined as 
\begin{equation} \label{eq:SNperzAndell}
    \left(\frac{S}{N}\right)^{z_i}_\ell = \sqrt{\sum_{XY} \left[\frac{\Delta C_\ell^{XY,z_i}}{\sigma_\ell^{XY,z_i}}\right]^2 },
\end{equation}
where $\Delta C_\ell^{XY,z_i} = C_\ell^{XY,z_i} - \tilde{C}_\ell^{XY}$ is the difference between the lensed and unlensed spectra, and the sum is taken over $XY \in \{TT,TE,EE,BB,\kappa\kappa\}$. 

We then consider wavenumber $k$ bins, uniformly spaced in $\log_{10}(k)$. For a given wavenumber bin~$j$ ranging from $k_{j,\mathrm{min}}$ to $k_{j,\mathrm{max}}$ and centered at $k_j$, we use the approximation $k \approx (\ell + 1/2) / \chi(z)$ to find the multipole range that corresponds to that $j$~bin for each redshift bin~$i$.  Thus we calculate $\ell_\mathrm{min}(z_i,k_j) = \chi(z_i) k_{j,\mathrm{min}} - 1/2$ and $\ell_\mathrm{max}(z_i,k_j) = \chi(z_i) k_{j,\mathrm{max}} - 1/2$. We then calculate the SNR within the $j^\mathrm{th}$ wavenumber bin and $i^\mathrm{th}$ redshift bin as 
\begin{equation} \label{eq:SNperzk}
    \left(\frac{S}{N}\right)^{z_i}_{k_j} = \sqrt{\sum_{\ell=\ell_\mathrm{min}(z_i,k_j)}^{\ell_\mathrm{max}(z_i,k_j)} \left[\left(\frac{S}{N}\right)^{z_i}_\ell\right]^2},
\end{equation}
adding the SNRs in quadrature. We sum over redshift bins to estimate the total SNR within the $j^\mathrm{th}$ wavenumber bin, by
\begin{equation} \label{eq:SNperk}
    \left(\frac{S}{N}\right)_{k_j} = \sqrt{\sum_i \left[ \left(\frac{S}{N}\right)^{z_i}_{k_j}\right]^2}.
\end{equation}

While Eq.~\ref{eq:SNperk} underestimates the noise by neglecting the off-diagonal elements of the covariance matrix, it also underestimates the signal; when summing over redshift bins by adding in quadrature, the lensing cross-terms between redshift bins are neglected in the squared signal (see Appendix~\ref{sec:SNRapprox}). These two effects coincidentally nearly cancel, yielding a reasonable approximation to the SNR. For example, using Eq.~\ref{eq:SNR} to calculate the lensing SNR from all the spectra yields 1947 (see Table~\ref{tab:lensing_snr} in Section~\ref{sec:results}), while using  Eq.~\ref{eq:SNperzAndell} and summing over all redshifts (to $z=1100$) and multipoles gives 1798, which is only slightly lower.\footnote{When using only $z \in [0,6]$ as shown in Figure~\ref{fig:SNR}, Eq.~\ref{eq:SNR} yields 1632, while  Eq.~\ref{eq:SNperzAndell} yields 1585 after summing over redshifts and multipoles, as shown in Table~\ref{tab:lensing_snr_approx} in Appendix~\ref{sec:SNRapprox}.}

\subsection{Including non-CDM Models}
\label{sec:wdm-baryons}

\begin{figure}[t]
    \centering
    \includegraphics[width=\columnwidth,height= 7.5cm]{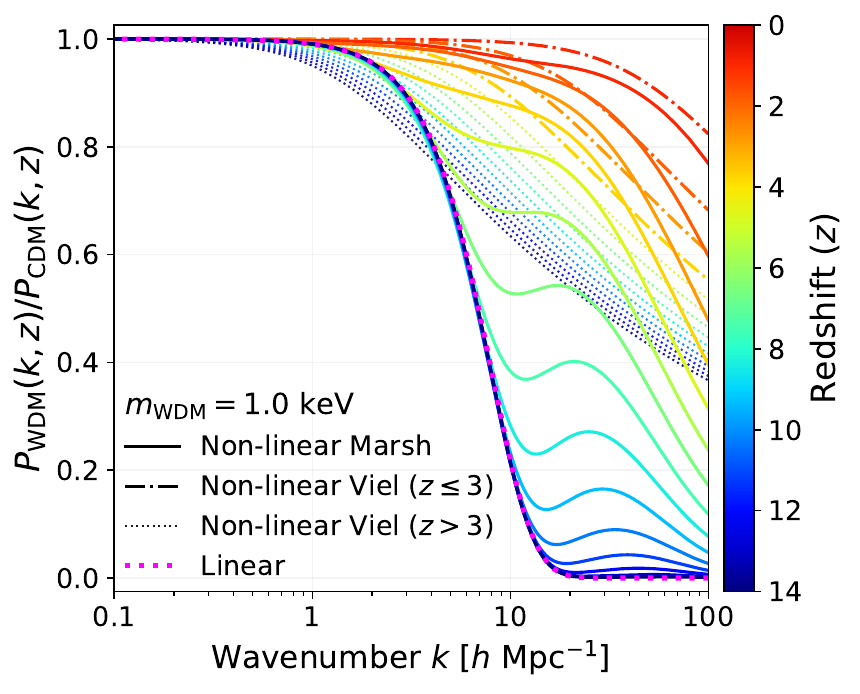}
    \caption{We show transfer functions, defined by Eq.~\ref{eq:transfer}, for a 1~keV warm dark matter (WDM) model. The solid curves show the transfer function for the non-linear matter power spectrum, calculated from the Warm\&Fuzzy code~\protect{\cite{Marsh2016}} at 15 evenly-spaced redshifts in the range $z \in [0,~14]$. At redshifts $z \gtrsim 14$, the non-linear transfer function converges to the transfer function for the linear matter power spectrum, given by Eq.~\ref{eq:lineartransfer}~\protect{\cite{Viel2005}} and shown as the magenta dotted curve. In this work, we use the WDM transfer function from Warm\&Fuzzy for redshifts $z \in [0,~14]$, and the linear transfer function at higher redshifts. For comparison, we also show the non-linear transfer function from~\protect{\cite{Viel2012}}, given by Eq.~\ref{eq:WDMtransfer}, for redshifts in the range $z \in [0,3]$ (dashed-dotted curves), which is the regime found to have a good match to simulations~\protect{\cite{Viel2012}}. We also show as small-dotted curves an extension of this model to $z=14$, and note that it does not converge to the linear transfer function at high $z$ (nor was it expected to be accurate in this regime).  We emphasize that we do not advocate for any particular model in this work, and acknowledge the complexity of modeling  non-linear dynamics and matching to simulations.}
    \label{fig:WDMtransfer}
\end{figure}

Models that are not purely CDM, but that have varying properties of the dark matter particle or include baryonic effects, yield non-linear matter power spectra that differ from that of CDM on small scales.  We include this effect by calculating the non-linear transfer function given by Eq.~\ref{eq:transfer}. While our formalism (and the public code we provide) allow for any non-linear transfer function, there are relatively few provided in the literature for alternate dark matter models. 

A non-linear transfer function for warm dark matter (WDM) is provided by~\cite{Viel2012} as
\begin{equation} \label{eq:WDMtransfer}
        T^2_\mathrm{non-lin}(k,z) = \left[1 + (\alpha k)^{\nu l}\right]^{-s / \nu},
\end{equation}
where $\nu = 3$, $l = 0.6$, $s = 0.4$, and $\alpha$ is given by
\begin{equation} \label{eq:nonlinearWDMalpha}
        \alpha(m_\mathrm{WDM},z) = 0.0476 \left(\frac{1~\mathrm{keV}}{m_\mathrm{WDM}}\right)^{1.85} \left(\frac{1+z}{2}\right)^{1.3}. 
\end{equation}
However,~\cite{Viel2012} find this fitting function to be accurate only for redshifts  $z < 3$ from comparison with simulations.  

Since we need to integrate over the matter power spectrum all the way back to recombination in Eq.~\ref{eq:ClPhiPhi}, we need a transfer function that extends to higher redshifts. Thus we calculate a non-linear transfer function from the Warm\&Fuzzy code~\cite{Marsh2016}, which extends to all redshifts. We obtain $P_\mathrm{m}^\mathrm{CDM}(k,z)$ and $P_\mathrm{m}^\mathrm{WDM}(k,z)$ for various WDM masses ranging from 1~to 10~keV keeping the other cosmological parameters fixed.\footnote{When calculating the WDM transfer function with the Warm\&Fuzzy code, we fix the cosmological parameters to $\Omega_\mathrm{m} = 0.31377$, $\Omega_\mathrm{b} = 0.04930$, $h = 0.6736$, $n_\mathrm{s} = 0.9649$ and $\sigma_8 = 0.811$, from~\cite{planck18params} and also given in Table~\ref{tab:fisher}.} We then apply the computed transfer function to the power spectrum $P_\Psi(k,z)$ from CAMB to obtain $C_\ell^{\kappa\kappa}$ from Eq.~\ref{eq:ClPhiPhi}, and to the kSZ power spectrum, described below in Section~\ref{sec:kSZ}.

We show the transfer function from~\cite{Marsh2016} for a WDM mass of 1~keV as the solid curves in Figure~\ref{fig:WDMtransfer} for different redshifts. As noted by~\cite{Marsh2016}, at high redshifts (e.g.,~$z>14$) the non-linear transfer function approaches the transfer function for the WDM {\it{linear}} matter power spectrum from~\cite{Viel2005,Bode2000}, for the range of $k$-modes considered in this work.  We show this linear transfer function as the magenta dotted curve in Figure~\ref{fig:WDMtransfer}, which is given by the fitting function
\begin{equation} \label{eq:lineartransfer}
    T^2_\mathrm{lin}(k,z) = \left[1 + (\alpha_\mathrm{lin} k)^{2\nu_\mathrm{lin}}\right]^{-10/\nu_\mathrm{lin}},
\end{equation}
from~\cite{Viel2005,Bode2000}, where $\nu_\mathrm{lin} = 1.12$ and $\alpha_\mathrm{lin}$ is given by 
\begin{equation} \label{eq:linearWDMalpha}
    \alpha_\mathrm{lin}(m_\mathrm{WDM}) = 0.049 \left(\frac{1~\mathrm{keV}}{m_\mathrm{WDM}}\right)^{1.11} \left(\frac{\Omega_\mathrm{WDM}}{0.25}\right)^{0.11} \left(\frac{0.7}{h}\right)^{1.22}.
\end{equation}
Here $\alpha_\mathrm{lin}$ is in units of $h^{-1}$~Mpc, $k$ is in units of $h$~Mpc$^{-1}$, $m_\mathrm{WDM}$ is the WDM mass in units of keV, $\Omega_\mathrm{WDM}$ is the WDM density parameter today, and $h$ is the reduced Hubble constant. Thus, we calculate the non-linear transfer function at 15 evenly-spaced redshifts in the range $z \in [0,~14]$ using the Warm\&Fuzzy code, and use the linear fit from Eq.~\ref{eq:lineartransfer} at higher redshifts. We show this in Figure~\ref{fig:WDMtransfer} for a 1~keV WDM mass, and also compare to the non-linear transfer function of~\cite{Viel2012} given by Eq.~\ref{eq:WDMtransfer}. We show as dashed-dotted curves the non-linear transfer function of Eq.~\ref{eq:WDMtransfer} for $z\leq3$, where~\cite{Viel2012} claims agreement with simulations at the 2\% level.  We also show as faint dotted curves this transfer function for $3 < z \leq 14$, which does not converge to the linear transfer function of Eq.~\ref{eq:lineartransfer}, but is also not expected to be accurate in this regime. 

In general, the two transfer functions are in rough agreement for $z<3$.  Figure 13 in~\cite{Schneider2012}, shows the non-linear transfer function from simulations at $z=0$ for a 1~keV WDM model, and compares that to the transfer function of~\cite{Viel2012}. From simulations,~\cite{Schneider2012} find a 2\% suppression in power at $k = 10 h$~Mpc$^{-1}$, whereas the transfer function from~\cite{Viel2012} predicts 0.7\% and the one from~\cite{Marsh2016} predicts 3.7\%.  We therefore emphasize how sensitive these transfer functions are to details of modelling.  For example, for 5~keV WDM, at $z=0.5$ and $k = 10 h$~Mpc$^{-1}$, the suppression predicted by~\cite{Viel2012} is 0.01\% and by~\cite{Marsh2016} is 0.1\%.  However,~\cite{Viel2012} only claims a 2\% agreement with simulations.  

\begin{figure}[t]
    \includegraphics[width=\columnwidth]{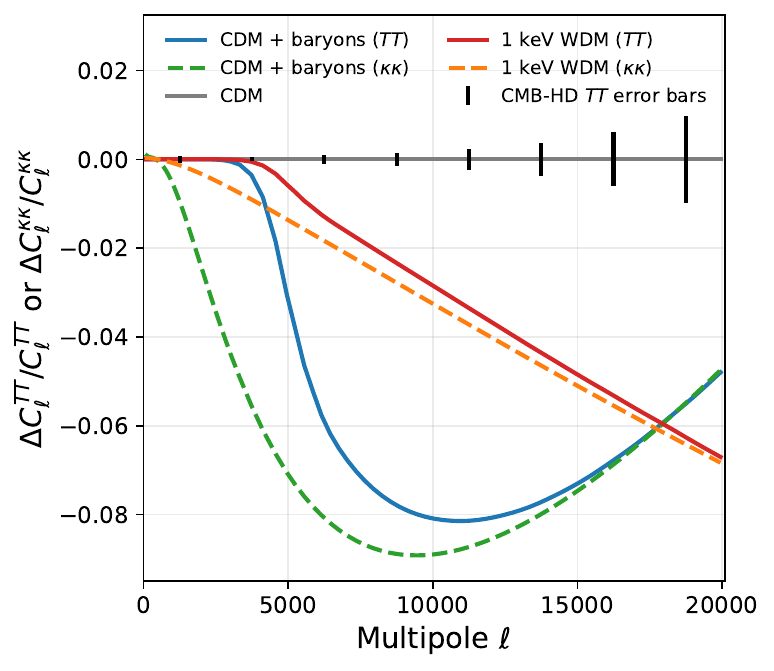}
    \caption{We show the fractional change in the power spectra, $\Delta C_\ell / C_\ell$, for a 1~keV WDM model versus a CDM-only model (red and orange) and for a CDM+baryons model versus a CDM-only model (blue and green). The spectra shown are the CMB temperature spectrum ($C_{\ell}^{TT}$, shown as solid curves) and the CMB lensing spectrum ($C_{\ell}^{\kappa\kappa}$, shown as dashed curves). The $C_{\ell}^{TT}$ spectra only include the effect of CMB lensing, not the kSZ effect; however, the matter power spectrum would show a similar behavior via the kSZ spectrum. We also plot the CMB-HD error bars on $\Delta C_\ell^{TT} / C_\ell^{TT}$, which we place on the CDM-only model. While both WDM and CDM+baryons models result in a suppression of power relative to a CDM-only model, the shape of the suppression differs for the two models, allowing CMB-HD to distinguish between them. } 
    \label{fig:TT_kk_ratio}
\end{figure}

We view the transfer functions discussed above as illustrative of the effects of alternate dark matter models, as opposed to definitive, and all WDM constraints presented below should be viewed with the understanding that they can change with updated transfer functions.  We also make public the code used for the calculations throughout this work\footnote{\url{https://github.com/CMB-HD/hdPk}}, which can be easily modified to use another non-linear transfer function in place of the one described above.

To include the effects of baryonic feedback on the non-linear matter distribution, we use the single-parameter baryonic feedback model of HMCode2020~\cite{Mead2020} within CAMB to calculate the non-linear power spectrum, $P_\Psi^\mathrm{CDM+baryons}(k,z)$, directly (instead of applying a transfer function).  This single-parameter model connects the various feedback parameters of hydrodynamic simulations to a single ``AGN temperature'' parameter, ${\rm{log_{10}}} (T_{\rm{AGN}}/{\rm{K}})$, which characterizes the overall strength of the feedback in the simulations.  While in general baryonic physics models can be quite complicated, a number of publications have suggested that the main effects can be characterized by a few parameters, or even as suggested by~\cite{Mead2020}, by one effective parameter via which several others can be characterized (see, e.g.~Table 5 in~\cite{Mead2020}).   We then follow the procedure described in Section~\ref{sec:signal} to calculate the power spectra with this single-parameter model.

We show in  Figure~\ref{fig:TT_kk_ratio} the change in the $TT$ or lensing $\kappa\kappa$ power spectrum from a 1~keV WDM or a CDM+baryonic feedback model, relative to a CDM-only model. For the $TT$ spectrum, we only show the change due to the difference in the lensing effect; the change in the kSZ effect would show a similar behavior. 
We see both WDM and CDM+baryonic feedback models result in a suppression of power relative to a CDM-only model, however, the shape of the suppression differs considerably between the two models.  A wide range of baryonic feedback models and simulations that extend beyond $k \sim 10$~$h$/Mpc, such as those considered in~\cite{Mead2020,Schneider2018,Giri2021,Salcido2023,Schaller2024}, seem to share a similar suppression at moderate scales followed by an enhancement at smaller scales (``U-shape'').  A WDM model, on the other hand, follows a scale-dependent suppression that increases steadily towards smaller scales, which is not easily mimicked by baryonic feedback.  Therefore, while the details of each kind of effect can be more complicated, Figure~\ref{fig:TT_kk_ratio} is still likely representative of CMB-HD's ability to distinguish between the effects of baryonic feedback and alternative dark matter.

\subsection{Modelling the kinetic Sunyaev-Zel'dovich Effect}
\label{sec:kSZ}

Since the kSZ effect is frequency-independent, it is harder to separate from the lensed CMB.  Thus we include the kSZ power spectrum in our analysis, and assume that it will need to be fit for along with the other cosmological parameters.\footnote{Parameters for frequency-dependent residual foregrounds will also need to be fit for in a full power spectrum analysis, however, we assume with the seven frequency channels of CMB-HD and all the resulting spectra and cross-spectra between frequency channels, that those other parameters will be well constrained and not degenerate with the $\Lambda $CDM parameters, as demonstrated by current CMB multi-frequency analyses~\cite{Planck2019Spectra, Choi2020,Balkenhol2022SPTspectra} (see, e.g.,~Fig.~7 in~\cite{han20delensing}.)}  We do not include measurements of the kSZ trispectrum in this work, but note that such measurements are an additional way to separate the kSZ signal from the lensed CMB, and would yield additional constraining power~\cite{Smith2016, Raghunathan2024}.

\begin{figure}[t]
    \centering
    \includegraphics[width=\columnwidth]{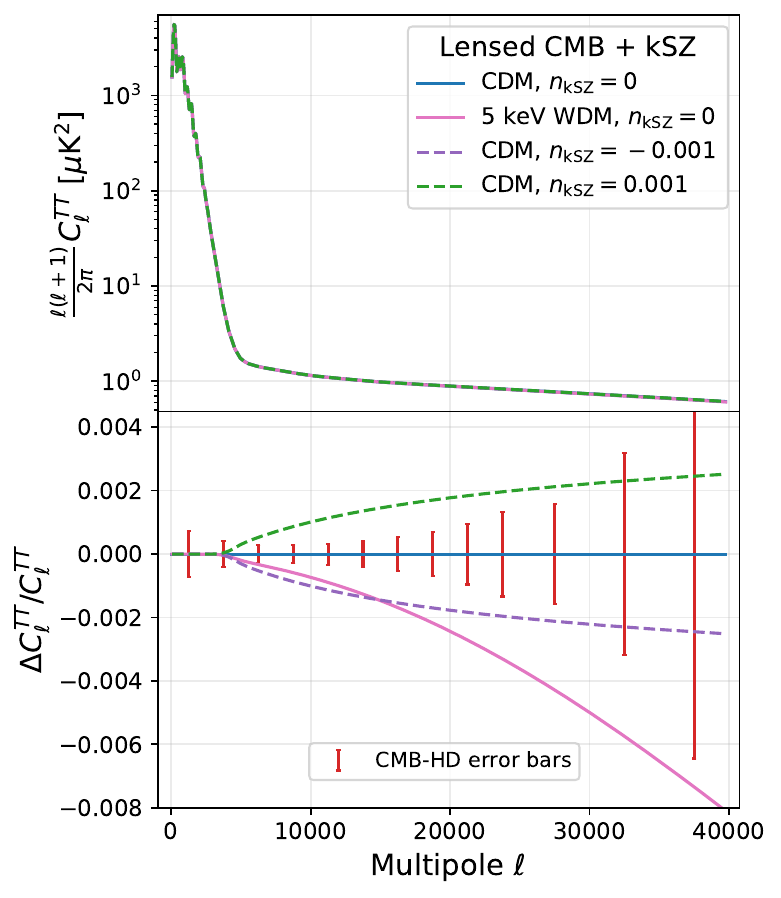}
    \caption{{\it{Top panel}}: We show the CMB temperature power spectrum, $C_\ell^{TT}$, including both the effects of lensing and the kSZ power spectra, calculated as described in Sections~\ref{sec:signal},~\ref{sec:wdm-baryons}, and~\ref{sec:kSZ}. We show $C_\ell^{TT}$ with our fiducial kSZ parameters ($A_\mathrm{kSZ}=1$ and $n_\mathrm{kSZ}=0$) in a fiducial CDM-only model (blue solid curve) and in a 5~keV WDM model (pink solid curve).  We also show $C_\ell^{TT}$ for a CDM-only model when we vary the kSZ slope (i.e.~$n_\mathrm{kSZ}=\pm 0.001$; green and purple dashed curves). The different $C_\ell^{TT}$ are indistinguishable by eye. {\it{Bottom panel}}: We show the fractional change in the $C_\ell^{TT}$ spectra described above, relative to the fiducial model. We see that changing the slope of the kSZ power spectrum in a CDM-only model does not reproduce the suppression due to WDM. We also show the forecasted CMB-HD error bars on $\Delta C_\ell^{TT} / C_\ell^{TT}$ (in red). } 
    \label{fig:ksz_ratio}
\end{figure}

To model the kSZ power spectrum, we use a template for the late-time kSZ power spectrum from~\cite{Battaglia2010}\footnote{The results are similar if we instead use the late-time kSZ template from~\cite{Omori2022}.} and for the reionization contribution from~\cite{Smith2018,Park2013}.  We extend these templates to $\ell_\mathrm{max} = 40,000$ using a linear fit to each template in the tail end of their existing range.  We follow the approach of~\cite{Dunkley2013} by normalizing the full kSZ spectrum $D_\ell^\mathrm{kSZ} \equiv \ell(\ell+1) C_\ell^\mathrm{kSZ} / 2\pi$ to equal 1~$\mu$K$^2$ at $\ell = 3000$.  Denoting the normalized template by $C_\ell^{\mathrm{kSZ},0}$, we model the kSZ power spectrum in a CDM-only model by, 
    \begin{equation} \label{eq:kSZpower}
        C_\ell^\mathrm{kSZ;CDM} = \left(\frac{\ell}{\ell_0}\right)^{n_\mathrm{kSZ}} A_\mathrm{kSZ} C_\ell^{\mathrm{kSZ},0},
    \end{equation}
where $\ell_0 = 3000$.  This model has two free parameters: $A_\mathrm{kSZ}$, which is the amplitude of the kSZ power spectrum at $\ell_0 = 3000$, and $n_\mathrm{kSZ}$, which is its slope.  We use fiducial values of $A_\mathrm{kSZ}$~=~1 and $n_\mathrm{kSZ}$~=~0 in this work, and allow these parameters to be free in the parameter analysis described below. Fitting $A_\mathrm{kSZ}$ with a similar kSZ template in the ACT DR4 power spectrum analysis (keeping $n_\mathrm{kSZ}$ fixed to our fiducial),~\cite{Choi2020} found $A_\mathrm{kSZ} < 1.8$ at 95\% CL.  We assume that measurements of the thermal Sunyaev-Zel'dovich (tSZ) power spectrum will constrain the large-scale shape of the kSZ power spectrum and $\ell_0$ pivot, since tSZ and kSZ trace the same large-scale structure; thus we do not free $\ell_0$.  

While the model we adopt here for the kSZ power spectrum with two free parameters is more flexible than that adopted in current CMB analyses, which usually just free $A_\mathrm{kSZ}$~\cite{Planck2019Spectra, Choi2020,Balkenhol2022SPTspectra}, it may still not be enough freedom to model the kSZ power spectrum in the uncharted small-scale regime that CMB-HD targets.  However, we note that we will have three additional handles on the small-scale shape of the kSZ power spectrum from measurements of 1)~the tSZ power spectrum, 2)~the kSZ trispectrum, and 3)~the kSZ cross-correlation with spectroscopic galaxy surveys. In addition, in this work we are interested in distinguishing $A_\mathrm{kSZ}$ and $n_\mathrm{kSZ}$ from $m_\mathrm{WDM}$, the mass of a WDM particle, and to aid in this we note that WDM needs to impact the lensing signal in $TT, TE, EE, BB$, and $\kappa \kappa$ as well as the kSZ power spectrum in a consistent fashion, whereas $A_\mathrm{kSZ}$ and $n_\mathrm{kSZ}$ only impact the kSZ power spectrum.

To model the kSZ power spectrum for WDM, we approximate the suppression of the kSZ power spectrum by applying the WDM transfer function defined in Eq.~\ref{eq:transfer} and discussed in Section~\ref{sec:wdm-baryons} to the kSZ power. We evaluate this transfer function at a fixed redshift $z_0 = 0.5$ to simplify the analysis, assuming the late-time kSZ dominates the kSZ spectrum.  We relate the comoving wavenumber $k$ to a multipole $\ell$ with the approximation $k \approx (\ell + 1/2) / \chi(z_0)$. For each WDM mass we define $T^2_\ell = T^2_\mathrm{non-lin}\bigl(k \approx (\ell + 1/2) / \chi(z_0)\bigr)$ and obtain the kSZ power spectrum in a WDM-only model by, 
    \begin{equation} \label{eq:kSZpowerWDM}
        C_\ell^\mathrm{kSZ;WDM} = T^2_\ell C_\ell^{\mathrm{kSZ;CDM}}.
    \end{equation}
We note that assuming all the kSZ signal is from the relatively low-redshift of $z_0 = 0.5$ is conservative.  The kSZ signal from higher redshifts, such as the contribution from reionization, will experience a larger suppression in power due to WDM, as can be seen from Figure~\ref{fig:WDMtransfer}.

We do not model the impact of baryonic feedback for the kSZ power spectrum, even though the change in the matter power spectrum due to baryonic effects will impact it.  This again is a choice to simplify the analysis, and is made because baryonic feedback already has such a large impact on the lensing signal in the modest multipole range of $\ell \sim 10,000$ where CMB-HD measurements are very constraining, as shown in Figure~\ref{fig:TT_kk_ratio}.  Thus, baryonic feedback parameters are already constrained well enough to not be degenerate with other parameters (see Section~\ref{sec:params} below).  In addition, tSZ power spectrum measurements will further inform the level of baryonic feedback.  However, fully modelling the kSZ effect, incorporating both a redshift-dependent non-CDM transfer function and baryonic effects, is one area where the modelling presented here could be extended.  

We show in the top panel of Figure~\ref{fig:ksz_ratio} the CMB temperature power spectrum, $C_\ell^{TT}$, including both the effect of lensing and the kSZ power spectrum.  We show this with the fiducial kSZ spectrum given by Eq.~\ref{eq:kSZpower} for a CDM-only model with the kSZ parameters $n_\mathrm{kSZ} = 0$ and $A_\mathrm{kSZ}=1$ (blue solid curve). We also show this for kSZ spectra with $n_\mathrm{kSZ} = \pm 0.001$ (purple and green dashed curves). We compare this to $C_\ell^{TT}$ for a 5~keV WDM-only model with the kSZ spectrum given by Eq.~\ref{eq:kSZpowerWDM} and with fiducial kSZ parameters ($n_\mathrm{kSZ} = 0$ and $A_\mathrm{kSZ}=1$); the WDM model modifies both the lensing and kSZ in this $C_\ell^{TT}$ (pink solid curve).  Since these spectra are indistinguishable by eye, in the bottom panel we show the fractional difference between these different $C_\ell^{TT}$ and our fiducial $C_\ell^{TT}$, along with the forecasted CMB-HD error bars on $\Delta C_\ell^{TT} / C_\ell^{TT}$ (in red).  We see that the suppression due to WDM cannot be easily mimicked by varying the slope of the kSZ power spectrum.

\subsection{Calculating the Fisher Matrix}
\label{sec:fisher}

We construct a Fisher matrix to forecast parameter constraints for CMB-HD. While we do not use a Markov chain Monte Carlo method (MCMC) in this work, which could be more accurate but also more time consuming, in~\cite{HDparams} we verified that our Fisher forecasts were consistent with the results obtained from an MCMC method. To calculate the Fisher matrix, we vary the parameters discussed below. \\

For the CDM-only model, we vary the six $\Lambda$CDM parameters: the physical density parameters for baryons, $\Omega_\mathrm{b} h^2$, and for cold dark matter (CDM), $\Omega_\mathrm{c} h^2$; the amplitude of the comoving primordial curvature power spectrum via the parameter $\ln\left(10^{10} A_\mathrm{s}\right)$, and the scalar spectral index $n_\mathrm{s}$, defined at the pivot scale $k_0 = 0.05$~Mpc$^{-1}$; the optical depth  to reionization, $\tau$; and the cosmoMC\footnote{\url{https://cosmologist.info/cosmomc/readme.html}} approximation to the angular sound horizon at recombination, $100 \theta_\mathrm{MC}$. We also vary the effective number of relativistic species $N_\mathrm{eff}$, the sum of the neutrino masses $\sum m_\nu$, and the amplitude  and slope of the kSZ power spectrum, $A_\mathrm{kSZ}$ and $n_\mathrm{kSZ}$ (see Section~\ref{sec:kSZ}). We refer to this as a $\Lambda$CDM + $N_\mathrm{eff}$ + $\sum m_\nu$ + $A_\mathrm{kSZ}$ + $n_\mathrm{kSZ}$ model. We adopt the \textit{Planck} baseline cosmological parameter constraints as our fiducial model~\cite{planck18params}. 

When we consider the case where all of the dark matter is warm, rather than cold, (the WDM-only model), we vary the ten parameters listed above, now with the CDM density parameter $\Omega_\mathrm{c} h^2$ replaced by the WDM density parameter $\Omega_\mathrm{w} h^2$, and additionally vary the mass of the WDM particle, $m_\mathrm{WDM}$. We refer to this eleven-parameter model as a $\Lambda$WDM + $N_\mathrm{eff}$ + $\sum m_\nu$ + $m_\mathrm{WDM}$ + $A_\mathrm{kSZ}$ + $n_\mathrm{kSZ}$ model.  In the figures and tables in Section~\ref{sec:results}, we use $\Omega_\mathrm{d} h^2$ as a general dark matter density parameter, to indicate either $\Omega_\mathrm{c} h^2$ or $\Omega_\mathrm{w} h^2$.

In both of the cases above, we also consider the effects of baryonic feedback, and vary the feedback parameter $\log_{10}(T_\mathrm{AGN}/\mathrm{K})$ of~\cite{Mead2020}. For a model with baryonic feedback, we also include a 0.06\% prior on $\log_{10}(T_\mathrm{AGN}/\mathrm{K})$, which is expected from tSZ measurements and was also included in~\cite{HDparams}.  In Table~\ref{tab:fisher}, we list the set of varied parameters, their fiducial values, and the step sizes used in the Fisher matrix calculation described below. All forecasts include a prior on the optical depth of $\sigma(\tau) = 0.007$ from \textit{Planck}~\cite{planck18params}.

\begin{table}[t]
    \begin{center}
    \begin{tabular}{l@{\hskip 1.5em} c@{\hskip 1.5em} c }
      \toprule
      \toprule
      Parameter & Fiducial value & Step size 
      \\
      \midrule
      $\Omega_\mathrm{b} h^2$\dotfill & $0.02237$ & 1\% 
      \\
      $\Omega_\mathrm{d} h^2$\dotfill & $0.1200$ & 1\%
      \\
      $\ln(10^{10} A_\mathrm{s})$\dotfill & $3.044$ & 0.3\%\footnote{The step size of 0.3\% for $\ln(10^{10} A_\mathrm{s})$ corresponds to an approximate step size of 1\% on $A_\mathrm{s}$.} 
      \\
      $n_\mathrm{s}$\dotfill & $0.9649$ & 1\%
      \\
      $\tau$\dotfill & $0.0544$ & 5\%
      \\
      $100 \theta_\mathrm{MC}$\dotfill & $1.04092$ & 1\%
      \\
      $N_\mathrm{eff}$\dotfill & $3.046$ & 5\%
      \\
      $\sum m_\nu$ [eV]\dotfill & $0.06$ & 10\%
      \\
      $\log_{10}\left(T_\mathrm{AGN}/\mathrm{K}\right)$\dotfill & 7.8 & 0.05 
      \\
      $m_\mathrm{WDM}$ [keV]\dotfill & [1,~10] & 10\%
      \\
      $A_\mathrm{kSZ}$\dotfill & 1 & 0.1
      \\
      $n_\mathrm{kSZ}$ \dotfill & 0 & 0.01
      \\
      \bottomrule
    \end{tabular}
    \caption{We list the cosmological parameters used in the Fisher forecasts, their fiducial values, and the step sizes used when varying the parameter values to compute the Fisher matrix. The fiducial values for the first six rows are taken from the~\textit{Planck} baseline cosmological parameters~\protect{\cite{planck18params}}, and their step sizes are taken as a fraction of their fiducial values. The fiducial value of the baryonic feedback parameter $\log_{10}\left(T_\mathrm{AGN}/\mathrm{K}\right)$ is taken from~\protect{\cite{Mead2020}}. We also list the range of values of the warm dark matter mass, $m_\mathrm{WDM}$, that we consider, and two parameters giving flexibility to the kSZ power spectrum ($A_\mathrm{kSZ}$ and $n_\mathrm{kSZ}$).  The step sizes for the baryonic feedback and kSZ parameters are absolute (i.e., not a relative fraction of their fiducial value). Note that we use $\Omega_\mathrm{d} h^2$ to refer to the physical dark matter density parameter, for either cold or warm dark matter.}
    \label{tab:fisher}
    \end{center}
\end{table}

To form a Fisher matrix for the CMB or BAO data, we assume a Gaussian likelihood function $\mathcal{L}(\hat{\vec{d}} | \vec{\theta})$ for the data vector $\hat{\vec{d}}$ given a set of theoretical model parameters~$\vec{\theta}$, 
\begin{equation}
        - 2 \ln \mathcal{L}\left(\hat{\vec{d}} | \vec{\theta}\right) = \sum_{ij} \Delta d_i(\vec{\theta}) \mathbb{C}^{-1}_{ij} \Delta d_j(\vec{\theta}),
\end{equation}
where $\Delta d_i(\vec{\theta})$ is the $i^\mathrm{th}$ element of $\Delta \vec{d}(\vec{\theta}) = \hat{\vec{d}} - \vec{d}(\vec{\theta})$, $\vec{d}(\vec{\theta})$ is the theoretical model of the data as a function of the parameters $\vec{\theta}$, and $\mathbb{C}^{-1}_{ij}$ is the $(i,j)$ element of the inverse covariance matrix for the data.  

For CMB-HD, the data vector $\hat{\vec{d}}$ is the set of binned CMB spectra (which could be lensed or delensed and includes the kSZ spectra) and the CMB lensing power spectrum. The covariance matrix is the binned covariance given by Eqs.~\ref{eq:GaussCovCMB} to~\ref{eq:eq:NonGaussCovCMBxLensing}, and the sum is taken over the bin centers. Each binned power spectrum is given by
\begin{equation}
        C_{\ell_b}^{XY} = \sum_\ell M_{b\ell} C_\ell^{XY},
\end{equation}
and the binned blocks of the covariance matrix are given by 
\begin{equation}
        \mathbb{C}_{\ell_b \ell_b'}^{XY,WZ} = \sum_{\ell \ell'} M_{b\ell} \mathbb{C}_{\ell \ell'}^{XY,WZ} M_{b'\ell'}^T,
\end{equation}
where $M_{b\ell}$ is the binning matrix, $\ell_b$ represents the bin center of the $b^\mathrm{th}$ bin, and $XY$ and $WZ$ can each be one of $\{TT, TE, EE, BB, \kappa\kappa\}$. 
       
For DESI, the data vector is the set of distance ratio measurements $r_s/d_V(z_j)$, with a covariance matrix given by $\mathbb{C}_{ij} = \delta_{ij} \sigma_i^2$ as discussed in Section~\ref{sec:desi}, and the sum is taken over the redshifts $z_i$.

The Fisher matrix $F$ describes how the likelihood changes when the parameters $\vec{\theta}$ are varied. Assuming that the likelihood is maximized for some fiducial set of parameters $\vec{\theta}_0$ and Taylor expanding about this point, the elements of the Fisher matrix are given by
\begin{equation} \label{eq:fisher}
        \begin{split}
            F_{\alpha\beta} & \equiv \left.-\left\langle \frac{\partial^2 \ln\mathcal{L}}{\partial \theta_\alpha \partial \theta_\beta} \right\rangle \right|_{\vec{\theta}_0} \\ & = \sum_{ij} \left.\left(\frac{\partial d_i}{\partial \theta_\alpha} \mathbb{C}_{ij}^{-1} \frac{\partial d_j}{\partial \theta_\beta}\right)\right|_{\vec{\theta}_0}.
        \end{split}
\end{equation}
Here the indices $\alpha$ and $\beta$ correspond to two parameters in $\vec{\theta}$. We use CAMB to evaluate the derivatives numerically for the cosmological and baryonic feedback parameters (first nine rows of Table~\ref{tab:fisher}), varying each parameter up or down by its step size while holding the other parameters fixed. To calculate the derivatives for $A_\mathrm{kSZ}$ and $n_\mathrm{kSZ}$, we similarly vary these parameters by the step sizes given in Table~\ref{tab:fisher}. For the WDM models, we vary the WDM mass in the transfer function of Eq.~\ref{eq:transfer}; this transfer function adjusts both the lensing effect and the kSZ spectrum. We change our fiducial model for each WDM mass considered (in the range 1 to 10 keV), keeping all other parameters fixed to their fiducial values. We then calculate derivatives for $m_\mathrm{WDM}$ by varying $m_\mathrm{WDM}$ around that fiducial mass by 10\% of its value.\footnote{We have confirmed that the parameter constraints are stable when the step sizes are varied slightly. We find a $1\sigma$ uncertainty that is smaller than the step size for all parameters except $\tau$ and $\sum m_\nu$, and have verified the CDM + baryons Fisher errors are consistent with those from an MCMC in~\cite{HDparams}.}

The inverse of the Fisher matrix gives the covariance matrix for the parameters, such that  $\left(F^{-1}\right)_{\alpha\beta} = \mathrm{cov}\left(\theta_\alpha, \theta_\beta\right)$; the uncertainty $\sigma_\alpha$ on the parameter $\theta_\alpha$ is found from its variance, $\sigma_\alpha^2 = \left(F^{-1}\right)_{\alpha\alpha}$. We apply a prior from \textit{Planck}~\cite{planck18params} on the optical depth $\tau$ of $\sigma(\tau) = 0.007$ by adding its inverse variance $1 / \sigma^2(\tau)$ to the element $F_{\tau\tau}^\mathrm{CMB}$ of the CMB Fisher matrix.

The Fisher matrix for the combination of CMB-HD and DESI is formed by taking the sum of their Fisher matrices, $F = F^\mathrm{CMB} + F^\mathrm{BAO}$.  To calculate $F^\mathrm{CMB}$ for CMB-HD, we use Eq.~\ref{eq:fisher} to calculate a Fisher matrix from the $TT, TE, EE, BB, \kappa\kappa$ power spectra in the range $\ell \in (30, 20000)$ with the (binned) covariance matrix given by Eqs.~\ref{eq:GaussCovCMB} to~\ref{eq:eq:NonGaussCovCMBxLensing}. We then calculate a second Fisher matrix from the $TT$ power spectrum in the range $\ell \in (20000,40000)$, with a diagonal covariance matrix given by Eq.~\ref{eq:GaussCovCMB}. We take the sum of these two Fisher matrices to obtain $F^\mathrm{CMB}$.

\section{Results}
\label{sec:results}

\subsection{CMB-HD Lensing from $TT$, $TE$, $EE$, $BB$, and $\kappa\kappa$}
\label{sec:lensing-snr}

We show in Table~\ref{tab:lensing_snr} the SNR with which CMB-HD can measure the CMB lensing signal. As described in Section~\ref{sec:snr}, we use Eq.~\ref{eq:SNR} where $\Delta C_\ell$ is the difference between lensed and unlensed spectra. In the first five rows, we give the SNR for each of the lensed CMB spectra $XY \in \{TT, TE, EE, BB\}$ and the CMB lensing $\kappa\kappa$ spectrum, over the multipole range $\ell \in [30, 20000]$.  We find roughly equal contributions to the lensing SNR from the lensed $TT$ power spectrum and the lensing $\kappa\kappa$ power spectrum, with the SNR from $TT$ (SNR=1571) being slightly higher than from $\kappa\kappa$ (SNR=1287).  We also find considerable SNR from the lensed $BB$ power spectrum alone (SNR=493).  In the sixth row, we show the SNR for $TT$ in the range $\ell \in (20000, 40000]$ (SNR=79); while this is modest compared to the SNR from lower multipoles, it is still almost double the final {\it{Planck}} lensing SNR~\cite{Planck2018lensing}.   In the last row of Table~\ref{tab:lensing_snr}, we give the total SNR from all the spectra.  We account for correlations between the spectra by using the full $5 n_\mathrm{bin} \times 5 n_\mathrm{bin}$ covariance matrix, described in Section~\ref{sec:covmat}, to calculate the SNR from all spectra in the range $\ell \in [30, 20000]$.  We obtain our final lensing SNR by adding in quadrature this value to the SNR from $TT$ in the range $\ell \in (20000, 40000]$. We find a total lensing SNR of 1947.

\begin{table}[t]
    \begin{center}
    \begin{tabular}{l @{\hskip 1em}c}
      \toprule
      \toprule
      Spectra  & Lensing SNR
      \\
      \midrule
      $TT$-only $\in (30, 20000)$\dotfill & 1571
      \\
      $TE$-only $\in (30, 20000)$\dotfill & 161
      \\
      $EE$-only $\in (30, 20000)$\dotfill & 276
      \\
      $BB$-only $\in (30, 20000)$\dotfill & 493
      \\
      $\kappa\kappa$-only ~$\in (30, 20000)$\dotfill & 1287 
      \\   
      $TT$-only $\in (20000, 40000)$\dotfill & 79
      \\
      All $TT+TE+EE+BB+\kappa\kappa$\dotfill & \textbf{1947}
      \\
      \bottomrule
    \end{tabular}
    \caption{We list the total signal-to-noise ratio (SNR) with which CMB-HD can measure the CMB lensing signal in a CDM-only model. The first six rows list the SNR values obtained only from the given spectrum $XY \in \{TT, TE, EE, BB, \kappa\kappa\}$ in the given multipole range. We use Eq.~\ref{eq:SNR} to calculate the SNR, with $\Delta C_{\ell}$ being the difference between the lensed and unlensed power spectrum. For the first five rows, $\mathbb{C}^{XY,XY}$ is given by the covariance matrix including non-Gaussian terms described in Section~\ref{sec:methods}. For the sixth row, we approximate the small-scale TT covariance matrix as diagonal, given by Eq.~\ref{eq:GaussCovCMB}, due to the high noise level on these scales. The last row gives the total SNR from the combined spectra listed in the first six rows. We calculate the total SNR from the first five spectra in the range $\ell \in (30,20000)$ from Eq.~\ref{eq:SNR} using the full $5 \times 5$ block covariance matrix, and then combine this value with the $TT$ SNR from $\ell \in (20000, 40000)$ by summing the two values in quadrature.}
    \label{tab:lensing_snr}
    \end{center}
\end{table}

\begin{figure*}[t]
    \centering
    \includegraphics[width=0.81\textwidth]{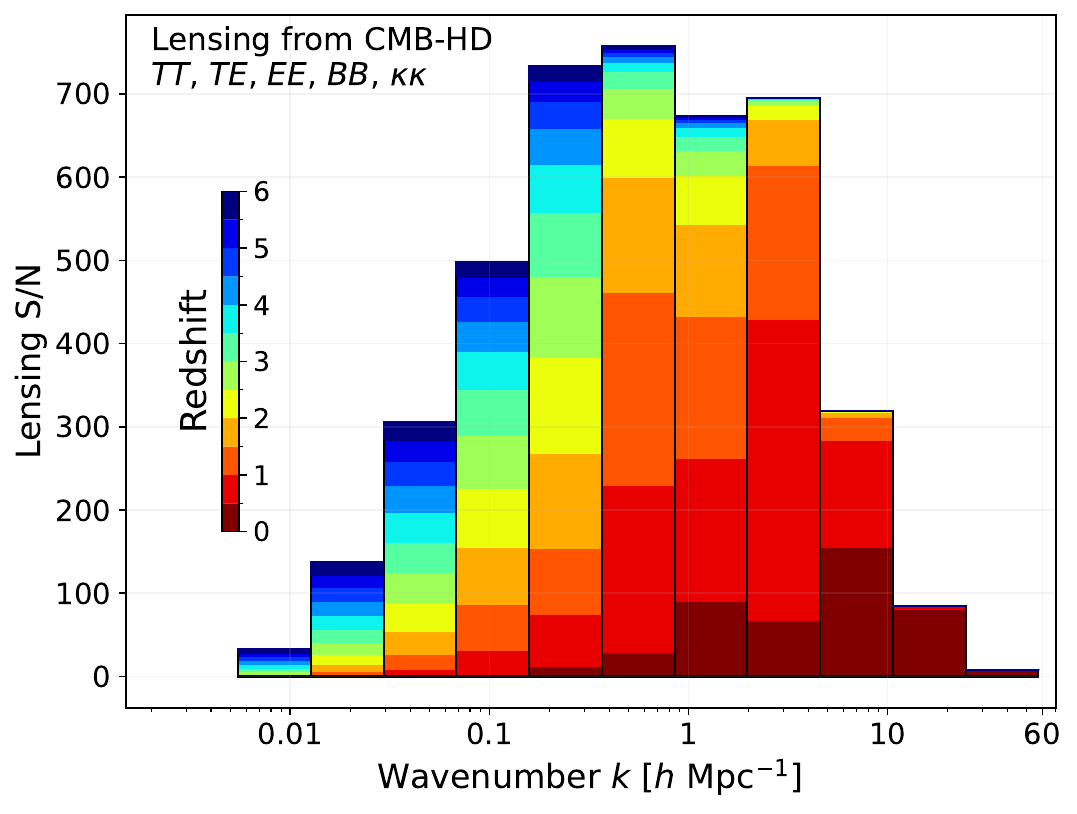}
    \includegraphics[width=0.79\textwidth]{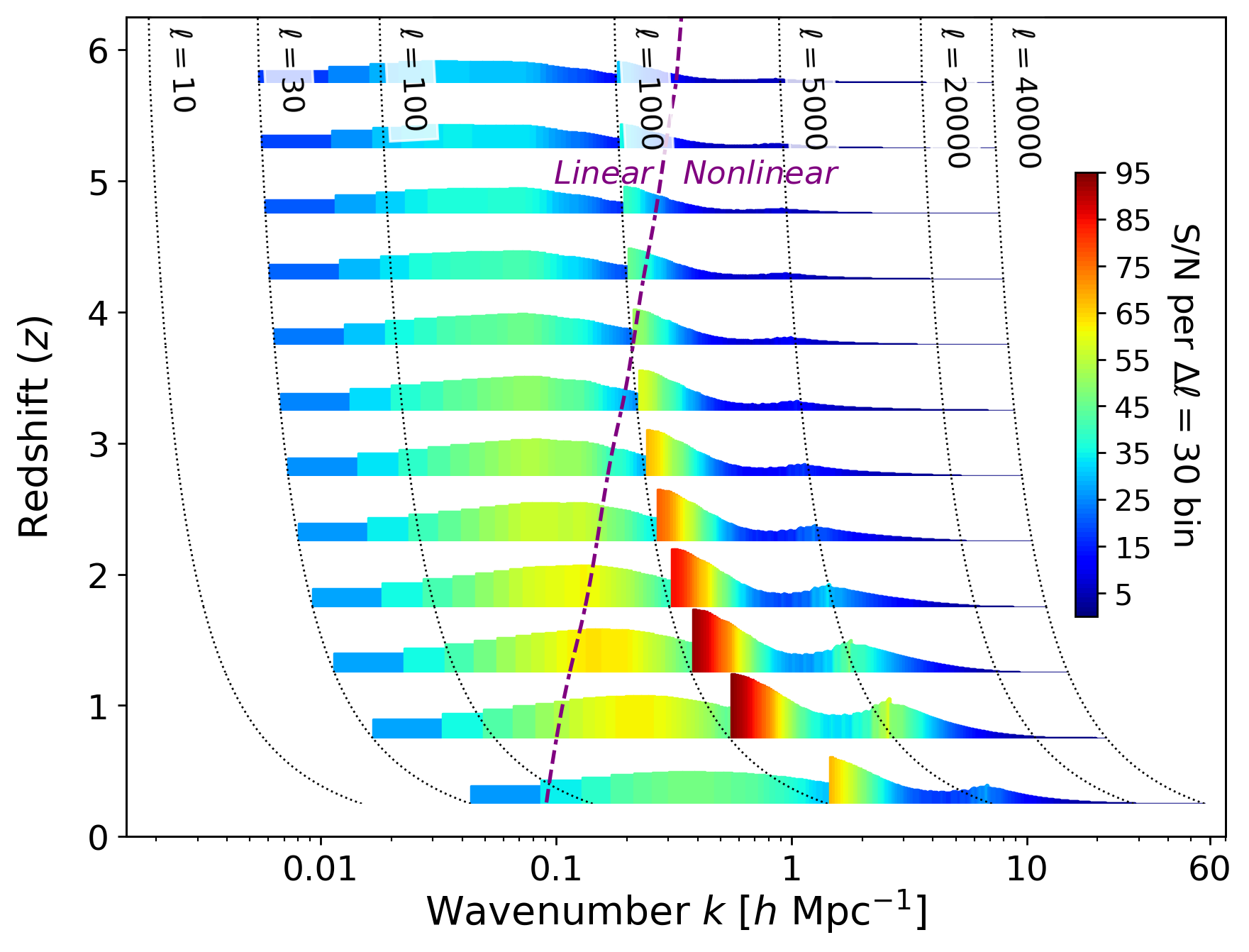}
    \caption{{\it{Top panel:}} We illustrate the approximate breakdown of the CMB-HD lensing SNR by comoving wavenumber and redshift for $z < 6$. (Note that the total SNR over all redshifts is 1947; for $z \leq 6$, the SNR is 1632.) {\it{Bottom panel:}} We further breakdown the lensing SNR per multipole bin.  The SNR increases at $\ell=1000$ due to the inclusion of the CMB-HD $BB$ spectrum which we assume has a minimum multipole of $\ell=1000$.  We also indicate a few multipoles (black dotted curves), and the linear and non-linear regimes (purple dashed curve). This figure was inspired by Figure 2 in~\protect{\cite{Green2021}}.}
    \label{fig:SNR}
\end{figure*}

In Figure~\ref{fig:SNR}, we illustrate which comoving scales and redshifts contribute to the CMB-HD lensing SNR given in the last row of Table~\ref{tab:lensing_snr}.  From Eq.~\ref{eq:ClPhiPhi}, we see that lensing from a range of comoving wavenumbers, $k$, integrated over all redshifts, contribute to lensing at a given multipole, $\ell$.  Thus we can only approximate the contribution to the lensing SNR for each redshift and wavenumber bin, as described in Section~\ref{sec:snr-perk} and Appendix~\ref{sec:SNRapprox}.

In the top panel of Figure~\ref{fig:SNR}, we plot a histogram of the lensing SNR per comoving wavenumber bin, $k$, and further break this down into the contribution from each redshift bin for $z \in [0, 6]$.  We see on large scales that the lensing SNR is derived from relatively high redshifts ($z>2$), while on smaller scales of $k < 1$~$h$~Mpc$^{-1}$ most of the SNR is from $z<2$. We find approximate lensing SNRs of SNR~$\approx 330$ for $k \gtrsim 5$~$h$~Mpc$^{-1}$ and SNR~$\approx 85$ for $k \gtrsim 10$~$h$~Mpc$^{-1}$.

In the bottom panel of Figure~\ref{fig:SNR}, we calculate the lensing SNR per multipole bin of width $\Delta \ell = 30$ and redshift bin, using Eq.~\ref{eq:SNperzAndell}.  We plot it at the wavenumber corresponding to the multipole bin center at that redshift. We indicate the SNR value by both the height of the bin and by its color, and also indicate a few multipoles as black dotted lines.  We show as the purple dashed line the transition between linear and non-linear scales at each redshift, which we derive by finding the wavenumber at which the linear matter power spectrum starts to be suppressed by more than 1\% relative to the non-linear matter power spectrum in a CDM-only model. 

From this figure, we see three peaks in the SNR at different scales.  This is from the contribution to the SNR from different spectra: $\kappa\kappa$ on the large scales (peak below $\ell = 1000$), $BB$ on intermediate scales (peak just above $\ell = 1000$, which is the minimum multipole where we assume CMB-HD will measure $BB$), and $TT$ on small scales (peak near $\ell = 5000$).  While the lensing measured from the $TT$ power spectrum on angular scales $\ell > 20,000$ provides a relatively small contribution to the total lensing SNR, as shown in Table~\ref{tab:lensing_snr}, it allows CMB-HD to probe the matter distribution out to scales corresponding to $k \approx 55$~$h$~Mpc$^{-1}$.  We also note that above $\ell>5000$ there are many multipole bins, so while each may not have high SNR, added together they are substantial, as indicated in the top panel of Figure~\ref{fig:SNR}.

In Figure~\ref{fig:Pk}, we illustrate how well CMB-HD lensing measurements can be used to measure the matter power spectrum.  We place the CMB-HD data points in Figure~\ref{fig:Pk} (shown in red) at the theoretical prediction (from CAMB) for the \textit{linear} matter power spectrum $P_\mathrm{m}^\mathrm{lin}(k)$.  We calculate lensing SNRs per $k$ bin the same way as in the top panel of Figure~\ref{fig:SNR}, this time doubling the number of $k$ bins. Since this SNR arises from the \textit{non-linear} matter distribution on a given scale, we equate $(S/N)_{k_j} = P_\mathrm{m}^\mathrm{non-lin}(k_j) / \sigma(P_\mathrm{m}^\mathrm{non-lin}(k_j))$. Following the approach of~\cite{Chabanier2019}, we use this to obtain the error bar on the linear matter power spectrum, 
\begin{equation} \label{eq:PkError}
    \begin{split}
            \sigma(P_\mathrm{m}^\mathrm{lin}(k_j))  & = P_\mathrm{m}^\mathrm{lin}(k_j) \left[\frac{\sigma(P_\mathrm{m}^\mathrm{non-lin}(k_j))}{ P_\mathrm{m}^\mathrm{non-lin}(k_j)}\right] \\ & = \frac{P_\mathrm{m}^\mathrm{lin}(k_j)}{(S/N)_{k_j}}.
    \end{split}
\end{equation} 
We note that in the non-linear HMCode~\cite{Mead2016} used in this work to obtain $P_\mathrm{m}^\mathrm{non-lin}(k)$, there is no mixing of wavenumbers between the linear and non-linear matter power spectra (the model assumes a one-to-one correspondence); thus Eq.~\ref{eq:PkError} is reasonable for illustrative purposes.

We follow the method of~\cite{Hlozek2012} to place the mass scale on the upper $x$-axis of Figure~\ref{fig:Pk}. We assume $k = 2\pi/\lambda$, where $\lambda$ is the diameter of a halo on scales corresponding to the comoving wavenumber $k$, and calculate the mass by 
\begin{equation} \label{eq:mass}
    M = \frac{4\pi}{3} \rho_\mathrm{m} \left(\frac{\pi}{k}\right)^3,
\end{equation}
where $\rho_\mathrm{m}$ is the total density in matter today.

The data points for the other experiments in Figure~\ref{fig:Pk} are taken from~\cite{Chabanier2019}\footnote{\url{https://github.com/marius311/mpk_compilation}}, which derives them following the method of~\cite{Tegmark2002}: the Lyman-$\alpha$ data (purple points) are derived from the 1D transmitted flux power spectrum~\cite{Chabanier2018} from the eBOSS DR14 release~\cite{eBOSSdr14}; the cosmic shear data (brown points) are derived from the DES Y1 measurements of the cosmic shear two-point correlation function~\cite{DESY1shear}; the galaxy clustering data (pink points) are derived from measurements of the halo power spectrum using luminous red galaxies (LRG) from the SDSS DR7 release~\cite{sdssDR7LRG}; the CMB data are derived from the \textit{Planck} 2018 temperature (blue points), polarization (yellow points), and lensing (green points) power spectra~\cite{Planck2018overview,Planck2018lensing}. We also include constraints derived from high-redshift measurements of the UV galaxy luminosity function from~\cite{Sabti2021}. The external data included in Figure~\ref{fig:Pk} does not represent the most recent measurements available 
(e.g.~\cite{Aiola2020,Choi2020,ACTdr6lensing,Pan2023,Dutcher2021,Balkenhol2022,Amon2021,Secco2021,Asgari2020,Li2023KiDS,Li2023HSC,Dalal2023,DES2021Clustering,Semenaite2021,Alam2016,deBelsunce2024,Gilman2021,Esteban2023}) since we mainly use the pre-computed data points derived by~\cite{Chabanier2019}. However, we make public the code used to generate Figure~\ref{fig:Pk}, and any other data sets can be readily added.

\subsection{Parameter Forecasts}
\label{sec:params}

We use the Fisher matrix method described in Section~\ref{sec:fisher}, to forecast parameter constraints from CMB-HD delensed $TT$, $TE$, $EE$, $BB$ and lensing $\kappa\kappa$ power spectra combinaed with DESI BAO data. In all cases we apply a prior of $\sigma(\tau) = 0.007$ from \textit{Planck}~\cite{planck18params}.

\begin{figure*}[t]
    \centering
    \includegraphics[width=\textwidth]{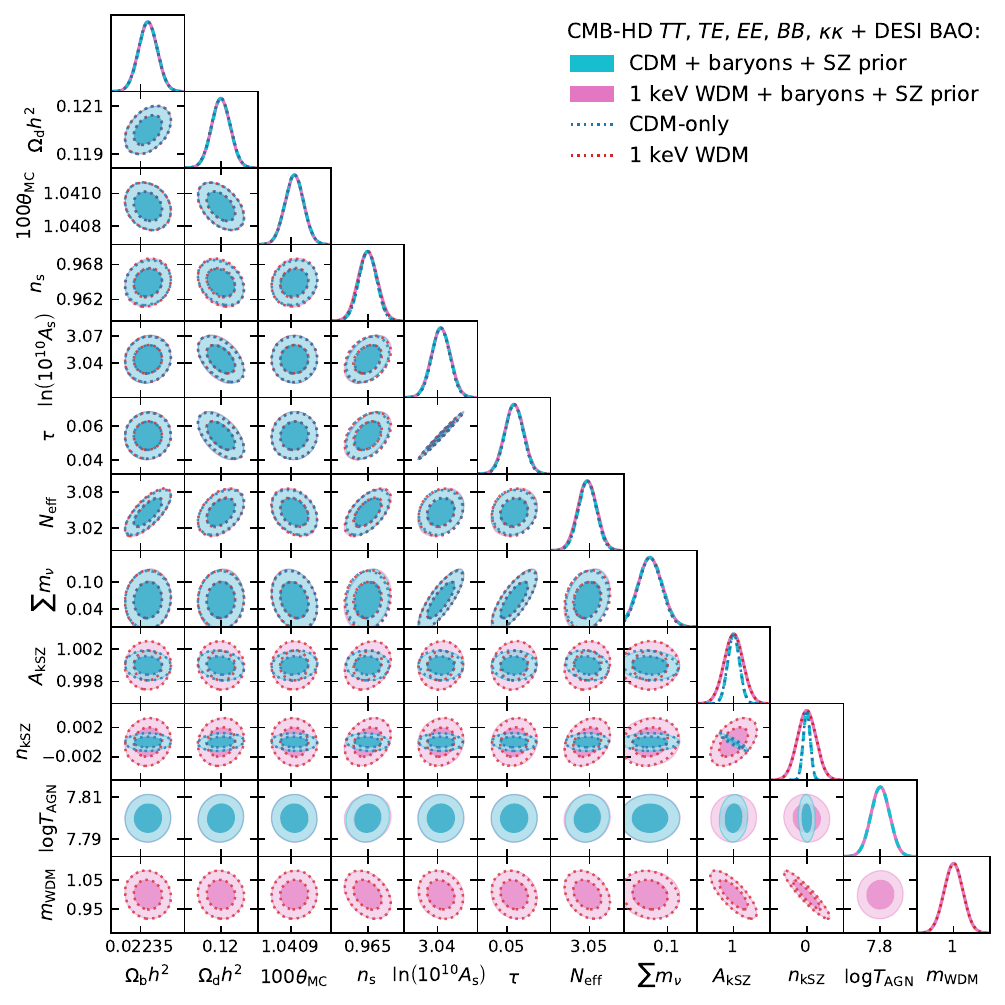}
    \caption{We show our forecasts including the full set of varied parameters. The blue dotted contours and curves show the parameter constraints in a $\Lambda$CDM + $N_\mathrm{eff}$ + $\sum m_\nu$ + $A_\mathrm{kSZ}$ + $n_\mathrm{kSZ}$ model assuming CDM. The red dotted contours and curves show the forecasts for a 1~keV $\Lambda$WDM + $N_\mathrm{eff}$ + $\sum m_\nu$ + $m_\mathrm{WDM}$ + $A_\mathrm{kSZ}$ + $n_\mathrm{kSZ}$ model.  We also show forecasts for these two models when including the effects of baryonic feedback (cyan contours and curves for CDM, pink contours and curves for WDM), with the strength of the feedback characterized by $\log_{10}(T_\mathrm{AGN}/\mathrm{K})$; we place a prior on this parameter from the anticipated cross-correlation of CMB-HD SZ and CMB lensing measurements as done in~\protect{\cite{HDparams}}. Table~\ref{tab:params} lists the $1\sigma$ parameter uncertainties corresponding to this figure.}
    \label{fig:Triangle}
\end{figure*}

\begin{figure}[t]
    \centering
    \includegraphics[width=\columnwidth]{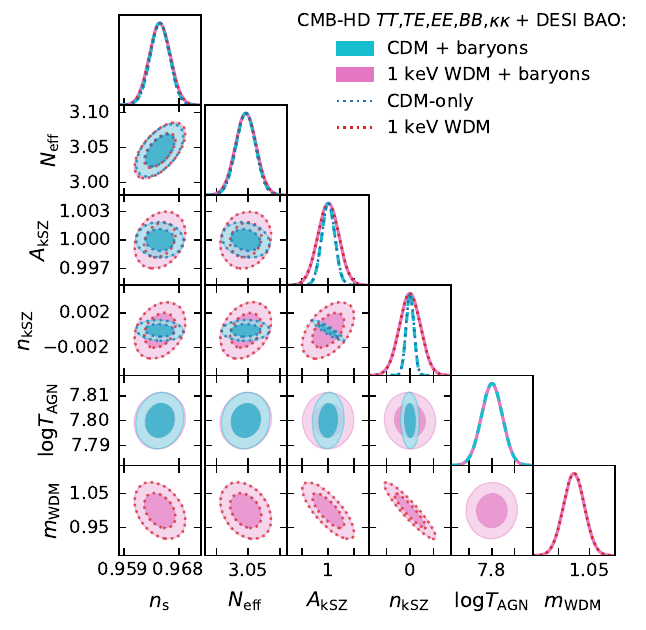}
    \caption{We show forecasts for a subset of the parameters varied in the Fisher analysis. For both CDM and WDM models, we marginalize over the baryonic feedback parameter, $\log_{10}(T_\mathrm{AGN}/\mathrm{K})$, with almost no increase in uncertainty on the other cosmological parameters (shown by the difference between dotted and solid contours of the same color).  For the WDM + baryons model, we find no degeneracy between $\log_{10}(T_\mathrm{AGN}/\mathrm{K})$ and the WDM mass, $m_\mathrm{WDM}$, indicating that they can be separately distinguished. }
    \label{fig:TriangleSmall}
\end{figure}

In Figures~\ref{fig:Triangle} and~\ref{fig:TriangleSmall} we show forecasts for a 1~keV WDM model, varying the full set of  parameters shown in Table~\ref{tab:fisher}.  Figure~\ref{fig:TriangleSmall} shows a subset of these parameters, highlighting negligible degeneracy between $m_\mathrm{WDM}$ and $\log_{10}\left(T_\mathrm{AGN}/\mathrm{K}\right)$, and a slight degeneracy between $m_\mathrm{WDM}$, $n_\mathrm{s}$, and $N_\mathrm{eff}$.  We see $m_\mathrm{WDM}$ is most degenerate with $A_\mathrm{kSZ}$, $n_\mathrm{kSZ}$.

In Table~\ref{tab:params}, we compare parameter forecasts for a CDM model, with and without baryonic feedback, and a 1~keV WDM model with baryonic feedback.  We find that including baryonic feedback plus an SZ prior has negligible impact on parameter constraints.  For the WDM model, we find a small increase in $\sigma(n_s)$, and significant increases in $\sigma(A_\mathrm{kSZ})$ and $\sigma(n_\mathrm{kSZ})$ compared to the CDM case, since these parameters are most degenerate with $m_\mathrm{WDM}$.

%

%\clearpage

\begin{table*}[t]
    \begin{center}
    \begin{tabular}{l@{\hskip 3em}  l@{\hskip 3em}  l@{\hskip 3em} l}
      \toprule
      \toprule
      \multirow{3}{*}{Parameter} & \multicolumn{3}{c}{$1\sigma$ Error from CMB-HD Delensed $TT$, $TE$, $EE$, $BB$, and $\kappa\kappa$ + DESI BAO}
      \\
      \cmidrule(){2-4}
       & CDM &   CDM + baryons + SZ prior  & 1 keV WDM  + baryons + SZ prior
      \\
      \midrule
        $\Omega_\mathrm{b} h^2$\dotfill & 0.000025   & ~~~~~~~~~~~~~~ 0.000025 & ~~~~~~~~~~~ 0.000025
        \\
        $\Omega_\mathrm{d} h^2$\dotfill & 0.00041  & ~~~~~~~~~~~~~~ 0.00041 & ~~~~~~~~~~~  0.00041
        \\
        $100\theta_\mathrm{MC}$\dotfill & 0.000060   & ~~~~~~~~~~~~~~ 0.000060 & ~~~~~~~~~~~  0.000060
        \\
        $\ln \left(10^{10} A_\mathrm{s}\right)$\dotfill & 0.010   & ~~~~~~~~~~~~~~ 0.011 & ~~~~~~~~~~~   0.011
        \\
        $n_\mathrm{s}$\dotfill & 0.0016   & ~~~~~~~~~~~~~~ 0.0016 & ~~~~~~~~~~~   0.0017
        \\
        $\tau$\dotfill & 0.0055  & ~~~~~~~~~~~~~~  0.0058 & ~~~~~~~~~~~  0.0058
        \\
        $N_\mathrm{eff}$\dotfill & 0.015  & ~~~~~~~~~~~~~~ 0.016 & ~~~~~~~~~~~  0.016
        \\
        $\sum m_\nu~$[eV]\dotfill & 0.026  & ~~~~~~~~~~~~~~  0.028 & ~~~~~~~~~~~  0.028
        \\
        $A_\mathrm{kSZ}$\dotfill & 0.00076 & ~~~~~~~~~~~~~~ 0.00075 & ~~~~~~~~~~~  0.0012
        \\
        $n_\mathrm{kSZ}$\dotfill & 0.00049 &  ~~~~~~~~~~~~~~ 0.00049 & ~~~~~~~~~~~  0.0013
        \\
        $\log_{10}(T_\mathrm{AGN}/\mathrm{K})$\dotfill & --- &  ~~~~~~~~~~~~~~ 0.0046 & ~~~~~~~~~~~   0.0046
        \\
        $m_\mathrm{WDM}~$ [keV] & ---  &   ~~~~~~~~~~~~~~ --- & ~~~~~~~~~~~  0.034
      \\
      \bottomrule
    \end{tabular}
    \caption{Parameter forecasts from the combination of CMB-HD delensed $TT$, $TE$, $EE$, $BB$ and lensing $\kappa\kappa$ power spectra with DESI BAO data. We consider two dark matter models: one in which all the dark matter is cold (columns 2 and 3), and one in which all the dark matter is warm with a mass of 1~keV (column 4). We list the parameters in the first column, and the 1$\sigma$ uncertainty on each parameter in the following columns; a blank value indicates the parameter is not included in the model. For CDM, we consider a model without baryonic feedback (column 2) and including baryonic feedback, applying an SZ prior on the baryonic feedback parameter, $\log_{10}(T_\mathrm{AGN}/\mathrm{K})$ (column 3). We use the same baryonic feedback model and prior for the 1~keV WDM model.  In all cases, we apply a prior on the optical depth, $\tau$, of $\sigma(\tau) = 0.007$ from \textit{Planck}~\protect{\cite{planck18params}}. We find that including baryonic feedback plus an SZ prior has negligible impact on parameter constraints.  For the WDM model, we find a small increase in $\sigma(n_s)$, and significant increases in $\sigma(A_\mathrm{kSZ})$ and $\sigma(n_\mathrm{kSZ})$ compared to the CDM case, since these parameters are most degenerate with $m_\mathrm{WDM}$.  We see that $\sigma(n_s)$ and $\sigma({N_{\rm{eff}}})$ are slightly larger than found in~\protect{\cite{HDparams}} due to the addition of the free kSZ parameters.}
    \label{tab:params}
    \end{center}
\end{table*}

\begin{table*}[t]
    \begin{center}
    \begin{tabular}{l@{\hskip 2em} c@{\hskip 1.5em} c@{\hskip 1.5em} c@{\hskip 1.5em} c@{\hskip 1.5em} c}
      \toprule
      \toprule
       & \multicolumn{5}{c}{$\sigma(m_\mathrm{WDM})$ from CMB-HD + DESI BAO}
      \\
      \cmidrule(){2-6}
      $m_\mathrm{WDM}$ & $m_\mathrm{WDM}$ only free        & $m_\mathrm{WDM}$ + $A_\mathrm{kSZ}$ + $n_\mathrm{kSZ}$ &   + $\Lambda$WDM  &  + $N_\mathrm{eff}$ + $\sum m_\nu$      & + baryons + SZ prior
      \\
      \midrule
      1 keV  & 0.0023 ($434.8\sigma$) & 0.026 ($38.5\sigma$) & 0.032 ($31.2\sigma$) & 0.034  ($29.4\sigma$) & \textbf{0.034} (\textbf{29.4}$\sigma$) 
      \\
      2 keV  & 0.013 ($153.8\sigma$)  & 0.20 ($10.0\sigma$)  & 0.22 ($9.1\sigma$)   & 0.22  ($9.1\sigma$)   & \textbf{0.22} (\textbf{9.1}$\sigma$) 
      \\
      3 keV  & 0.044 ($68.2\sigma$)   & 0.30 ($10.0\sigma$)  & 0.31 ($9.7\sigma$)   & 0.31  ($9.7\sigma$)   & \textbf{0.31} (\textbf{9.7}$\sigma$) 
      \\
      4 keV  & 0.11 ($36.4\sigma$)    & 0.56 ($7.1\sigma$)   & 0.57 ($7.0\sigma$)   & 0.58  ($6.9\sigma$)   & \textbf{0.57} (\textbf{7.0}$\sigma$) 
      \\
      5 keV  & 0.24 ($20.8\sigma$)    & 1.1 ($4.5\sigma$)    & 1.1 ($4.5\sigma$)    & 1.1 ($4.5\sigma$)     & \textbf{1.1} (\textbf{4.5}$\sigma$) 
      \\
      6 keV  & 0.45 ($13.3\sigma$)    & 1.9 ($3.2\sigma$)    & 1.9 ($3.2\sigma$)    & 1.9 ($3.2\sigma$)     & \textbf{1.9} (\textbf{3.2}$\sigma$) 
      \\
      7 keV  & 0.73 ($9.6\sigma$)     & 3.2 ($2.2\sigma$)    & 3.2 ($2.2\sigma$)    & 3.3 ($2.1\sigma$)     & \textbf{3.2} (\textbf{2.2}$\sigma$) 
      \\
      8 keV  & 1.2 ($6.7\sigma$)      & 4.8 ($1.7\sigma$)    & 4.9 ($1.6\sigma$)    & 5.0 ($1.6\sigma$)     & \textbf{4.9} (\textbf{1.6}$\sigma$) 
      \\
      9 keV  & 1.8 ($5.0\sigma$)      & 7.3 ($1.2\sigma$)    & 7.5 ($1.2\sigma$)    & 7.5 ($1.2\sigma$)     & \textbf{7.5} (\textbf{1.2}$\sigma$) 
      \\
      10 keV  & 2.3 ($4.3\sigma$)     & 11.0 ($0.9\sigma$)   & 11.0 ($0.9\sigma$)   & 11.0 ($0.9\sigma$)    & \textbf{11.0} (\textbf{0.9}$\sigma$) 
      \\
      \bottomrule
    \end{tabular}
    \caption{We show the forecasted constraints on the WDM mass, $m_\mathrm{WDM}$, from CMB-HD delensed $TT, TE, EE, BB$ and $\kappa\kappa$ power spectra plus DESI BAO.  We list both the forecasted $1\sigma$ uncertainty on $m_\mathrm{WDM}$ in units of keV and the significance with which CMB-HD can measure this $m_\mathrm{WDM}$ (number in parentheses). We show these constraints when only $m_\mathrm{WDM}$ is varied, when the kSZ parameters are also varied, when the six $\Lambda$WDM parameters are additionally varied, when $N_\mathrm{eff}$ and $\sum m_\nu$ is varied as well, and when the baryonic feedback parameter, $\log_{10}(T_\mathrm{AGN}/\mathrm{K})$, is also allowed to be free.    We see that only varying $m_\mathrm{WDM}$ results in much tighter constraints than when also varying the kSZ parameters.  Varying all the other parameters has minimal impact.  CMB-HD can measure 1~keV WDM at about $30\sigma$ significance and rule out about 7~keV WDM at $2\sigma$. 
    } 
    \label{tab:WDMerror}
    \end{center}
\end{table*}

In Table~\ref{tab:WDMerror}, we list the forecasted $1\sigma$ uncertainty on the WDM mass, $\sigma(m_\mathrm{WDM})$, for fiducial WDM models with masses ranging from 1 to 10 keV.  We also quantify the significance with which we can measure a non-zero WDM mass by taking the ratio $m_\mathrm{WDM} / \sigma(m_\mathrm{WDM})$.  For each WDM mass considered, we calculate this uncertainty for five different cases: 1) freeing only $m_\mathrm{WDM}$ with all other parameters fixed to their fiducial values; 2) additionally freeing the kSZ parameters, $A_\mathrm{kSZ}$ and $n_\mathrm{kSZ}$; 3) additionally freeing the six $\Lambda$WDM parameters (but fixing $N_\mathrm{eff}$ and $\sum m_\nu$); 4) also freeing  $N_\mathrm{eff}$ and $\sum m_\nu$; and 5) also freeing the baryonic feedback parameter $\log_{10}\left(T_\mathrm{AGN}/\mathrm{K}\right)$.  As expected, our constraints are strongest when we only vary the WDM mass.  Varying the kSZ parameters $A_\mathrm{kSZ}$ and $n_\mathrm{kSZ}$ significantly degrades these constraints.  We find that varying the cosmological parameters and baryonic feedback minimally impacts constraints on the WDM mass once we allow flexibility in the shape of the kSZ spectrum. In the scenario where all parameters are freed, we find CDM-HD can measure 1~keV WDM mass with $30\sigma$ significance, and can rule out about 7~keV WDM at the 95\% confidence level.

\section{Discussion}
\label{sec:discussion}

In this work, we find that CMB-HD can measure gravitational lensing with a total significance of $1947\sigma$.  To put this in context, current best constraints on CMB lensing are $40\sigma$ from \textit{Planck}~\cite{Planck2018lensing} and $43\sigma$ from ACT~\cite{ACTdr6lensing}. This lensing measurement would span $0.005~h$/Mpc$~< k <~55~h$/Mpc, over four orders of magnitude in scale. The smallest scales probed by CMB-HD would correspond to halo masses of about $10^8 M_{\odot}$ today. 

We find that most of the lensing SNR comes from the $TT$ spectrum, although the $\kappa\kappa$ spectrum has a roughly equal contribution.  We find that the $BB$ power spectrum alone can measure CMB lensing with a significance of almost $500\sigma$.  On the smallest scales, most of the lensing SNR is from the $TT$ spectrum, which suggests a different focus for efforts to remove foreground contamination from lensing on these scales; it may be more important to lower the foreground contribution to the $TT$ spectrum or constrain this contribution well with multi-frequency observations, than to remove non-Gaussian foreground contributions from the lensing trispectrum ($\kappa\kappa$).  In addition, more optimal lensing estimators than used in this work for $\kappa\kappa$ have been proposed~\cite{Horowitz:2017iql,Schaan:2018tup,Hadzhiyska2019,Millea:2020iuw,Madhavacheril:2020ido,Legrand:2021qdu,Sailer:2022jwt,Chan2023,Legrand:2023jne}, and have the potential to yield higher lensing signal-to-noise ratios than forecasted in this work.

Since the kSZ signal is frequency-independent, it is harder to separate this foreground from the CMB than frequency-dependent foregrounds.  Thus, we include it in this work as part of the $TT$ spectrum.  We find that including the kSZ spectrum improves constraints on non-CDM models since the kSZ effect is a dominant signal at large multipoles ($\ell>5000$), and non-CDM models impact the matter power spectrum it traces.  Figure 3 of~\cite{Antypas2022} shows how the kSZ spectrum from CMB-HD can also constrain fractional amounts of ultralight dark matter, extending the work of~\cite{Farren2021}.

We also find that baryonic feedback effects change the shape of the matter power spectrum in ways that differ significantly from the change induced by non-CDM models.  While a more comprehensive or complex feedback model may alter the details of the suppression, a broad range of numerical simulations that include hydrodynamics suggest a relatively universal behavior~\cite{Mead2020,Schneider2018,Giri2021,Salcido2023,Schaller2024}, which is not easily reproduced by alternate dark matter models.  Since CMB-HD measures lensing accurately over a wide range of scales, it can readily detect this baryonic feedback effect at over $100\sigma$.  Thus we expect in general CMB-HD to overcome a significant challenge in efforts to measure non-CDM models, which is that one often cannot tell if a given small-scale deviation from CDM is due to an alternative dark matter model or baryonic feedback effects.  More refined and robust theory modeling of these effects would help to improve the accuracy of these forecasts and conclusions.

Current efforts to constrain non-CDM models have focused in large part on ruling out WDM.  These efforts use measurements of the Lyman-$\alpha$ forest, Milky Way satellite galaxies, strong lensing of quasars, and stellar streams~\cite{Baur2015,Garzilli2019,Gilman2019,Hsueh2019,Banik2019,Nadler2020,Villasenor2022,Irsic2023,Hooper2022,Keeley2024,Salucci2018}.  These constraints suggest roughly 5~keV WDM is ruled out at the 95\% confidence level, with some spread due to differing assumptions or techniques.  Here we present an independent and complimentary method with different systematics that can inform the particle properties of dark matter. An advantage of this technique is that the CMB lensing signal can be calculated theoretically from first principles, given the non-linear matter power spectrum.  In addition, the CMB lensing signal can be measured in multiple different ways, which themselves have different systematics, allowing for important cross checks.  While the kSZ power spectrum involves some astrophysics modelling, being sensitive to the product of the free electron density and velocity, there are a few different external handles on the kSZ effect from measurements of the tSZ power spectrum, the kSZ trispectrum, and cross-correlation with spectroscopic galaxy surveys. We can also see that limiting the flexibility in the kSZ power spectrum shape, with, for example, priors from external measurements, would improve WDM constraints considerably.  \\

Future CMB-HD measurements will not only be able to probe the early Universe via measurements of inflation and light relics, they will be a powerful and complementary probe of structure in the Universe down to sub-galactic scales and of dark matter particle properties.

\begin{acknowledgments}
The authors thank Boris Bolliet, Anthony Challinor, Francis-Yan Cyr-Racine, Rouven Essig, Gil Holder, Mathew Madhavacheril, David Marsh, Joel Meyers, Julian Munoz, Blake Sherwin, Anže Slosar, and Alexander van Engelen for useful discussions.  The authors thank Miriam Rothermel for testing the code and Jupyter notebooks made public with this paper, and Keyi Chen for an early version of the bottom panel of Figure~6. AM and NS acknowledge support from DOE award numbers DE-SC0020441 and DE-SC0025309 and the Stony Brook OVPR Seed Grant Program.  This research used resources of the National Energy Research Scientific Computing Center (NERSC), a U.S. Department of Energy Office of Science User Facility located at Lawrence Berkeley National Laboratory, operated under Contract No.~DE-AC02-05CH11231 using NERSC award HEP-ERCAPmp107.

\end{acknowledgments}

\appendix

\section{Extragalactic Foregrounds} \label{sec:extragalacticFGs}

As discussed in Section~\ref{sec:covmat}, our forecasts account for the residual extragalactic foreground signal that is expected to remain in the CMB-HD maps after foreground-subtraction procedures. Our extragalactic foreground model includes contributions from the thermal and kinetic Sunyaev-Zel’dovich effects (tSZ and kSZ, respectively), from the cosmic infrared background (CIB), and radio sources.

The CIB and radio galaxies appear as point sources in the maps, with a known shape corresponding to the instrument beam. Given the resolution and sensitivity of CMB-HD, as shown in Table~\ref{tab:ExpConfig}, we expect that most of the CIB and radio sources will be removed by resolving these sources down to a very low flux limit, measuring their fluxes, and subtracting them directly from the map. The advantage of this template subtraction approach, as opposed to masking sources, is that we do not cut holes in the map or disturb the underlying lensing signal. 

The CMB-HD single-frequency $5\sigma$ flux limits are given in Table~3 of~\cite{HDsnowmass}, calculated assuming white noise. Given the very high-resolution of CMB-HD, the addition of diffuse extragalactic components (such as the CMB, tSZ, and kSZ) has a negligible impact on these flux limits when using a matched filter to identify point sources using the beam shape; we have also verified this with simulations. In this work, we conservatively assume that radio and CIB galaxies will be identified using only a single frequency channel for each component (90 GHz for radio sources and 280 GHz for the CIB). In reality, one would actually use a multi-frequency matched filter, which would leverage the seven frequency channels of CMB-HD and the frequency dependence of these sources.

We assume that the 90~GHz channel will be used to find radio galaxies above 0.04~mJy with more than $5\sigma$ significance. Since the 90~and 150~GHz are relatively close in frequency, we expect that sources found in 90~GHz maps will also have well measured fluxes at 150~GHz; the 150~GHz $5\sigma$ flux limit is 0.05~mJy~\cite{HDsnowmass}, so this is a reasonable assumption. Assuming a spectral index of -0.8~\cite{Sehgal2010}, sources detected above 0.04~mJy at 90~GHz correspond to sources above 0.03~mJy at 150~GHz.  So we take 0.04~mJy and 0.03~mJy to be the radio source flux limits for 90 and 150~GHz, respectively.

We assume that CIB sources will be identified using the 280~GHz maps, where they are much brighter than their counterparts at 90~and 150~GHz. Thus we extrapolate the flux measured at 280 GHz to 90 and 150 GHz to properly subtract the sources at the lower frequencies. The $5\sigma$ flux limit at 280 GHz is 0.1~mJy~\cite{HDsnowmass}, but we allow a higher flux limit of 0.15~mJy to be more conservative. Assuming a spectral index of 2.6 for CIB sources~\cite{Sehgal2010} results in the identification of sources above 0.03~mJy at 150~GHz and 0.008~mJy at 90~GHz.  We note that there can be uncertainty in the CIB spectral index for a given source, which can lead to over- or under-subtraction of CIB sources at 90~and 150~GHz. Thus we incorporate about 5\% flux mis-subtraction corresponding to roughly 5\% spectral index uncertainty in our residual CIB foreground levels~\cite{han22}.\footnote{We expect precursor CMB experiments will have well characterized the CIB spectral index prior to CMB-HD measurements.}

We also assume that we can remove all tSZ clusters detected at $3\sigma$ by a multi-frequency matched filter. We use the cluster mass threshold for $5\sigma$ detected clusters forecast for CMB-HD by~\cite{Raghunathan2021} and shown in their Figure 3. We then convert this to a mass threshold for $3\sigma$ detected clusters by multiplying their mass threshold by 0.75, which assumes their Y-M scaling relation of $\textrm{Y} \propto M^{1.79}$.  We assume clusters above this mass threshold can be subtracted from CMB-HD maps.

The kSZ signal consists of two components, 1)~a contribution from the epoch of reionization and 2)~a late-time, lower-redshift, component (hereafter reionization kSZ and late-time kSZ, respectively)~\cite{Smith2018}. We include both of these components in this work as contaminants to the CMB power spectrum, and only the reionization part of the kSZ as a contaminant to the lensing power spectrum on small-scales. (We note that new lensing estimators being developed, e.g.~\cite{Chan2023}, have novel methods of separating the kSZ from lensing, and this can also be accomplished with kSZ trispectrum measurements~\cite{Smith2016}.)

We show in~\cite{han22} that inclusion of all the residual foregrounds described above does not result in appreciable bias to the lensing power spectrum.

\section{Construction of the Covariance Matrix} \label{sec:HDcovmatcalc}

As mentioned in Section~\ref{sec:CovmatConstruction}, we follow a similar approach to~\cite{HDparams} to calculate the CMB-HD covariance matrix for the lensed or delensed CMB and the CMB lensing spectra, which we describe in detail below. 

The blocks of the lensed and delensed covariance matrices are calculated analytically using the CLASS delens package~\cite{Hotinli2021} (unless otherwise specified). We denote each block by $\mathbb{C}_{\ell_1 \ell_2}^{XY,WZ}$, where $XY$ and $WZ$ can each be one of $TT$, $TE$, $EE$, $BB$, and $\kappa\kappa$.    This calculation accounts for Gaussian and non-Gaussian terms (i.e.,~diagonal and off-diagonal elements, respectively), including:

\begin{itemize}
    \item The Gaussian variance between different CMB spectra, given by
    \begin{equation} \label{eq:GaussCovCMB}
            \mathbb{C}_{\ell_1 \ell_2,\mathrm{G}}^{XY,WZ}  = \frac{\delta_{\ell_1 \ell_2} f_\mathrm{sky}^{-1}}{2 \ell_1 + 1} \left[C_{\ell_1}^{XW,\mathrm{tot}} C_{\ell_1}^{YZ,\mathrm{tot}} + C_{\ell_1}^{XZ,\mathrm{tot}} C_{\ell_1}^{YW,\mathrm{tot}}\right] ,
        \end{equation}
        where $C_{\ell}^{UV}$ refers to a lensed or delensed CMB power spectrum for $UV \in \{TT, TE, EE, BB\}$, and $C_{\ell}^{UV,\mathrm{tot}} = C_{\ell}^{UV} + N_{\ell}^{UV}$ is the sum of the signal and noise spectra (including residual foregrounds for the latter, as described above in Sections~\ref{sec:fg} and~\ref{sec:CMBnoise}). 

        \item The Gaussian variance of the CMB lensing power spectrum, given by 
        \begin{equation} \label{eq:GaussCovLensing}
            \mathbb{C}_{\ell_1 \ell_2,\mathrm{G}}^{\phi\phi,\phi\phi}  = \delta_{\ell_1 \ell_2} \frac{2 f_\mathrm{sky}^{-1}}{2 \ell_1 + 1} \left(C_{\ell_1}^{\phi\phi} + N_{\ell_1}^{\phi\phi}\right)^2 ,
        \end{equation}
        where $C_\ell^{\kappa\kappa} = [\ell(\ell+1)]^2 C_\ell^{\phi\phi}/4$, and the noise $N_\ell^{\kappa\kappa} = [\ell(\ell+1)]^2 N_\ell^{\phi\phi}/4$ on the CMB lensing power spectrum is described in Section~\ref{sec:LensingNoise}.

        \item The non-Gaussian lensing-induced covariances between different modes of the CMB spectra (e.g., $TT \times TT$, $TT \times EE$, etc.) given by
        \begin{equation} \label{eq:NonGaussCovCMBxCMB}
            \mathbb{C}_{\ell_1 \ell_2,\mathrm{NG}}^{XY,WZ}  = \sum_{\ell} \frac{\partial C_{\ell_1}^{XY}}{\partial C_\ell^{\phi\phi}} \frac{2 f_\mathrm{sky}^{-1}}{2 \ell + 1} \left(C_\ell^{\phi\phi}\right)^2 \frac{\partial C_{\ell_1}^{WZ}}{\partial C_\ell^{\phi\phi}},
        \end{equation}
        where $XY$ and $WZ$ are each one of $TT$, $TE$, $EE$. 
        
        \item The non-Gaussian lensing-induced covariances between $BB$ and any of the CMB spectra, given by
        \begin{equation} \label{eq:NonGaussCovCMBxBB}
            \begin{split}
                \mathbb{C}_{\ell_1 \ell_2,\mathrm{NG}}^{XY,BB}  & = (1 - \delta_{\ell_1 \ell_2}) \sum_\ell \left[\frac{\partial C_{\ell_1}^{XY}}{\partial C_\ell^{XY,\mathrm{u}}} \mathbb{C}_{\ell\ell,\mathrm{G}}^{XY,EE;\mathrm{u}} \frac{\partial C_{\ell_2}^{BB}}{\partial C_\ell^{EE,\mathrm{u}}}\right]  \\ & + \sum_{\ell} \frac{\partial C_{\ell_1}^{XY}}{\partial C_\ell^{\phi\phi}} \frac{2 f_\mathrm{sky}^{-1}}{2 \ell + 1} \left(C_\ell^{\phi\phi}\right)^2 \frac{\partial C_{\ell_1}^{BB}}{\partial C_\ell^{\phi\phi}},
            \end{split}
        \end{equation}
        where
        \begin{equation} 
            \mathbb{C}_{\ell \ell,\mathrm{G}}^{XY,WZ;\mathrm{u}}  = \frac{f_\mathrm{sky}^{-1}}{2 \ell + 1} \left[C_{\ell}^{XW,\mathrm{u}} C_{\ell}^{YZ,\mathrm{u}} + C_{\ell}^{XZ,\mathrm{u}} C_{\ell}^{YW,\mathrm{u}}\right],
        \end{equation}
        for $XY \in \{TT, TE, EE, BB\}$,        
        and $C_{\ell}^{UV,\mathrm{u}}$ is the unlensed CMB spectrum (or, if $UV = BB$, we replace it with the unlensed $EE$ spectrum, $C_{\ell}^{EE,\mathrm{u}}$).\footnote{In~\cite{HDparams} and~\cite{Hotinli2021},  Eq.~\ref{eq:NonGaussCovCMBxBB} was used for all the off-diagonal CMB $\times$ CMB covariances by setting $BB$ to any $WZ \in \{TT, TE, EE, BB\}$. However,~\cite{BenoitLevySmithHu2012} claim that, when $XY$ and $WZ$ are $\in \{TT, TE, EE\}$, then the first term in Eq.~\ref{eq:NonGaussCovCMBxBB} is cancelled by other higher-order corrections. This claim seems to be supported by unpublished preliminary tests with simulations we have seen. Thus we limit $WZ$ to only $BB$ in Eq.~\ref{eq:NonGaussCovCMBxBB} in this work.} 

        \item The non-Gaussian covariances between the CMB spectra and the CMB lensing spectrum,
        \begin{equation} \label{eq:eq:NonGaussCovCMBxLensing}
            \mathbb{C}_{\ell_1 \ell_2,\mathrm{NG}}^{\phi\phi, XY} =   \frac{2 f_\mathrm{sky}^{-1}}{2 \ell_1 + 1} \left(C_{\ell_1}^{\phi\phi}\right)^2 \frac{\partial C_{\ell_2}^{XY}}{\partial C_{\ell_1}^{\phi\phi}}.
        \end{equation}

        \item The non-Gaussian covariances of the CMB lensing spectrum, $\kappa\kappa \times \kappa\kappa$, for small scales ($\ell_1, \ell_2 > 5000$) given by the simulation-based covariance matrix from~\cite{han22}.  
    \end{itemize}

\noindent We additionally calculate the lensing covariance from modes that are larger than the survey size (i.e.~super-sample covariance (SSC))~\cite{Manzotti2014,Motloch2018}. This SSC term is given by~\cite{Manzotti2014,Motloch2018}
\begin{equation*}
    \mathbb{C}_{\ell_1 \ell_2,\mathrm{SSC}}^{XY,WZ} = \frac{\partial \ell_1^2 C_{\ell_1}^{XY}}{\partial \ln \ell_1} \frac{\sigma_\kappa^2}{\ell_1^2 \ell_2^2} \frac{\partial \ell_2^2 C_{\ell_2}^{WZ}}{\partial \ln \ell_2},
\end{equation*}
where $XY$ and $WZ$ are each one of $\{TT, TE, EE, BB\}$.  Here $\sigma_\kappa^2$ is the variance of the lensing convergence field within the observed patch of sky given by
\begin{equation*}
    \sigma_\kappa^2 = \frac{1}{A^2} \sum_{\ell m} \left|W_{\ell m}\right|^2 C_\ell^{\kappa\kappa},
\end{equation*}
where $W_{\ell m}$ is the harmonic transform of the window function, and $A$ is the sky area observed in radians. 
We find that including this SSC term in our covariance matrices does not change any parameter forecasts due to the large sky area of CMB-HD ($f_\mathrm{sky}=60\%$).  Thus, we do not include it in our final covariance matrices, as also done in~\cite{HDparams}.

The total covariance for the CMB $\times$ CMB blocks is then the sum of Eq.~\ref{eq:GaussCovCMB} and either~\ref{eq:NonGaussCovCMBxCMB} or~\ref{eq:NonGaussCovCMBxBB}, $\mathbb{C}_{\ell_1 \ell_2}^{XY,WZ} = \mathbb{C}_{\ell_1 \ell_2, \mathrm{G}}^{XY,WZ} + \mathbb{C}_{\ell_1 \ell_2,\mathrm{NG}}^{XY,WZ}$.  We assemble the full five-by-five block covariance matrix by joining the CMB$\times$CMB blocks with the CMB$\times$lensing (Eq.~\ref{eq:eq:NonGaussCovCMBxLensing}) and lensing$\times$lensing blocks  (Eq.~\ref{eq:GaussCovLensing} plus the simulation-based covariance matrix from~\cite{han22}).

We also compute a separate covariance matrix for the lensed and delensed $TT$ spectrum in the multipole range $\ell \in [20000, 40000]$. We assume this covariance matrix is diagonal, which is a good approximation given the large noise levels on these small scales.\footnote{We confirm that this is true by calculating the lensing SNR for $TT$ in the adjacent multipole range of $\ell \in [15000, 20000]$ using either the full covariance matrix (i.e.~including off-diagonal elements) or a diagonal-only covariance matrix, and find that the SNRs are the same in both cases.} We use Eq.~\ref{eq:GaussCovCMB} to calculate these diagonal elements.  For the SNRs presented in Section~\ref{sec:results}, we use this diagonal covariance matrix to calculate the SNRs for $TT$ in the range $20,000 < \ell < 40,000$, and add the resulting ratios in quadrature to the ratios calculated from all spectra in the range $30~<~\ell~<~20,000$. We describe in Section~\ref{sec:fisher} how we include the covariance matrix for $TT$ on scales $20,000 < \ell < 40,000$ in our parameter forecasts.

\section{Comparison of Analytic Spectra Calculations to CAMB} \label{sec:CAMBComparison}

While CAMB can calculate the CMB lensing power spectrum and the lensed and delensed CMB $TT$, $TE$, $EE$, and $BB$ power spectra in a CDM-only or CDM + baryons model, it cannot calculate this for an arbitrary non-CDM model.  Thus, to model the impact of non-CDM models that change the matter power spectrum, we need to calculate the modified $C_\ell^{\kappa\kappa}$ outside of CAMB, and then feed this $C_\ell^{\kappa\kappa}$ into CAMB to generate lensed and delensed CMB $TT, TE, EE$, and $BB$ spectra. Since we compare results from non-CDM models to those from CDM models, we calculate all of the CMB-HD power spectra in this way for consistency. This ensures that any difference between sets of power spectra is due to a difference in the model, rather than a difference in the method used to calculate the spectra. 

\begin{figure}[t]
    \centering
    \includegraphics[width=\columnwidth]{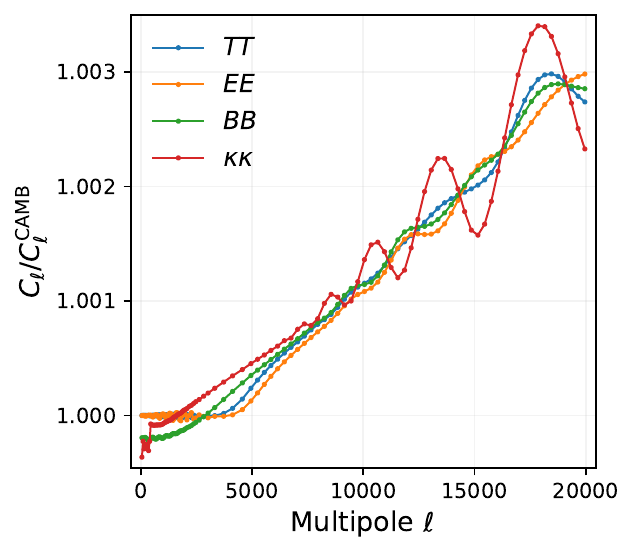}
    \caption{We show the ratio of delensed CMB $TT$, $EE$, $BB$ and CMB lensing $\kappa\kappa$ power spectra, for a CDM model with $C_\ell^{\kappa\kappa}$ computed using Eq.~\ref{eq:ClPhiPhi} versus obtained directly from CAMB. In both cases, the delensed CMB power spectra are calculated as described in Section~\ref{sec:signal}, using the corresponding $C_\ell^{\kappa\kappa}$. Our calculations are consistent with CAMB within 0.3\% up to $\ell = 20,000$ for all spectra, and within 0.8\% for $TT$ out to $\ell = 40,000$ (latter not shown).}
    \label{fig:SpectraComparisonCAMB}
\end{figure}

In Figure~\ref{fig:SpectraComparisonCAMB}, we compare the CMB lensing spectrum, $C_\ell^{\kappa\kappa}$, from Eq.~\ref{eq:ClPhiPhi} with the CAMB-only result for a CDM-only model to test the accuracy our calculation. In our calculation of Eq.~\ref{eq:ClPhiPhi}, we obtain $P_\Psi(k,z)$ from CAMB initially. We use identical cosmological and accuracy parameters in both cases (Eq.~\ref{eq:ClPhiPhi} versus CAMB only) and then take the ratio of their spectra (shown in red).  We also compare the delensed $TT$, $EE$, and $BB$ spectra obtained by passing the residual lensing power given by Eq.~\ref{eq:ResidualLensingPower} to CAMB, where the residual lensing is calculated using either our calculation of $C_\ell^{\kappa\kappa}$, or CAMB's internal calculation. We find that our calculation of the CMB and lensing power spectra agree with CAMB to within 0.3\% out to $\ell = 20,000$, which is the region that accounts for nearly all of the CMB-HD lensing SNR, as shown in Table~\ref{tab:lensing_snr}. Our calculation of $C_\ell^{TT}$ differs from that of CAMB by at most 0.8\% out to $\ell = 40,000$.

\section{Approximating the Lensing Signal-to-Noise Ratio per k Mode}\label{sec:SNRapprox}

Figure~\ref{fig:SNR} illustrates how a given comoving physical scale at a given redshift $z$ contributes to the total lensing SNR listed in Table~\ref{tab:lensing_snr}. This information is also used to visualize the CMB-HD constraints on the matter power spectrum shown in Fig.~\ref{fig:Pk}.  As discussed in Section~\ref{sec:snr-perk}, we can only approximate how much lensing comes from a specific comoving wavenumber ($k$) and redshift ($z$) since measurable angular multipoles ($\ell$) of lensing are drawn from an integral over $k$ and $z$.  The two approximations we make are 1)~assuming the covariance matrix is diagonal to obtain a lensing SNR value for each redshift and wavenumber bin, and 2)~summing the SNRs in quadrature over redshift bins to obtain the total lensing SNR per wavenumber bin.  The first approximation underestimates the noise by neglecting the off-diagonal elements of the covariance matrix.  The second approximation underestimates the signal by ignoring the lensing cross-terms between the redshift bins when adding SNRs from redshift bins in quadrature. These two effects nearly cancel, yielding a reasonable approximation, as will discuss show below. We note that this approximation procedure is only used to create Figures~\ref{fig:Pk} and~\ref{fig:SNR}; we use Eq.~\ref{eq:SNR} with the full covariance matrix described in Section~\ref{sec:covmat} to calculate the total lensing SNR values presented in Table~\ref{tab:lensing_snr}.

We begin with a set of theoretical lensed and unlensed CMB power spectra, $C_\ell^{XY}$ and $\tilde{C}_\ell^{XY}$, respectively, and with a set of CMB-HD noise curves $N_\ell^{XY}$, for $XY \in \{TT,TE,EE,BB,\kappa\kappa\}$. We first calculate a SNR per multipole, $(S/N)_\ell$. We do this by considering the unbinned version of Eq.~\ref{eq:SNR} for a single spectrum, $XY \in \{TT, TE, EE, BB, \kappa\kappa\}$:
    \begin{equation}
        \left(\frac{S}{N}\right)^{XY} = \sqrt{\sum_{\ell \ell'} (\Delta C_\ell^{XY}) (\mathbb{C}^{XY,XY})^{-1}_{\ell\ell'} (\Delta C_{\ell'}^{XY})},
    \end{equation}
where the sum is taken from $\ell_\mathrm{min}$ to $\ell_\mathrm{max}$. Assuming the covariance matrix is diagonal, $\mathbb{C}^{XY,XY}_{\ell\ell'} = \delta_{\ell\ell'} \left(\sigma_\ell^{XY}\right)^2$, this reduces to 
    \begin{equation}
        \left(\frac{S}{N}\right)^{XY} = \sqrt{\sum_{\ell} \left(\frac{\Delta C_\ell^{XY}}{\sigma_\ell^{XY}}\right)^2 },
    \end{equation}
which allows us to define a SNR per multipole $\ell$, 
    \begin{equation}
        \left(\frac{S}{N}\right)_\ell^{XY} = \frac{\left|\Delta C_\ell^{XY}\right|}{\sigma_\ell^{XY}}.
    \end{equation}
Approximating the full $5 \times 5$ block covariance matrix as diagonal, with blocks $\mathbb{C}_{\ell\ell'}^{XY,WZ} = \delta_{XY,WZ} \delta_{\ell\ell'} \left(\sigma_\ell^{XY}\right)^2$, yields a total SNR given by 
\begin{equation}
    \frac{S}{N} = \sqrt{\sum_{XY} \sum_{\ell} \left(\frac{\Delta C_\ell^{XY}}{\sigma_\ell^{XY}}\right)^2 } = \sqrt{\sum_{\ell} \sum_{XY} \left[\left(\frac{S}{N}\right)_\ell^{XY}\right]^2 }.
\end{equation}
The total lensing SNR per multipole $\ell$ is then 
\begin{equation}
    \left(\frac{S}{N}\right)_\ell = \sqrt{\sum_{XY} \left[\left(\frac{S}{N}\right)_\ell^{XY}\right]^2 }.
\end{equation}

Next we calculate an approximate lensing SNR per multipole and {\it{per redshift}}. We consider redshifts in the range $0 < z < 6$, using 12 redshift bins of width $\Delta z = 0.5$.  The contribution $C_\ell^{\phi\phi,z_i}$ to the CMB lensing power spectrum from the $i^\mathrm{th}$ redshift bin, ranging from $z_{i,\mathrm{min}}$ to $z_{i,\mathrm{max}}$ and centered at $z_i$, is calculated by integrating Eq.~\ref{eq:ClPhiPhi} from $\chi(z_{i,\mathrm{min}})$ to $\chi(z_{i,\mathrm{max}})$. This lensing power spectrum is then used to lens the unlensed CMB $TT$, $TE$, $EE$, and $BB$ power spectra as described in Section~\ref{sec:signal}, resulting in a set of five spectra $C_\ell^{XY,z_i}$ with a lensing signal arising only from $z_{i,\mathrm{min}} < z < z_{i,\mathrm{max}}$. We calculate an approximate covariance matrix for each redshift bin~$i$, again only including elements along the diagonal.  The diagonals of each covariance matrix are calculated for each multipole~$\ell$, so for the $i^\mathrm{th}$ redshift bin we define $\mathbb{C}^{XY,WZ;z_i}_{\ell \ell'} = \delta_{XY,WZ} \delta_{\ell \ell'} \left(\sigma_\ell^{XY,z_i}\right)^2$, where 
\begin{equation} \label{eq:KnoxForZbin}
    \begin{split}
        \left(\sigma_\ell^{XY,z_i}\right)^2 = \frac{f_\mathrm{sky}^{-1}}{(2\ell + 1)}  \Bigl[& \left(C_\ell^{XX,z_i} + N_\ell^{XX}\right) \left(C_\ell^{YY,z_i} + N_\ell^{YY}\right) \\ & + \left(C_\ell^{XY,z_i} + N_\ell^{XY}\right)^2\Bigr],
    \end{split}
\end{equation}
$C_\ell^{UV,z_i}$ is the power spectrum including lensing only from the $i^\mathrm{th}$ redshift bin, and $N_\ell^{UV}$ is the noise on the power spectrum, which we assume is the same for each redshift bin. (This is the same as Eq.~\ref{eq:GaussCovCMB} for $XY = WZ$, which results in Eq.~\ref{eq:GaussCovLensing} when $XY = \phi\phi$, but replacing each $C_\ell^\mathrm{tot} = C_\ell + N_\ell$ with $C_\ell^{z_i} + N_\ell$).  We then define the lensing SNR per multipole~$\ell$ in the $i^\mathrm{th}$ redshift bin as 
\begin{equation} \label{eq:SNRperZbinAndEll}
    \left(\frac{S}{N}\right)^{z_i}_\ell = \sqrt{\sum_{XY} \left[\left(\frac{S}{N}\right)_\ell^{XY,z_i}\right]^2 },
\end{equation}
where
\begin{equation}
    \left(\frac{S}{N}\right)_\ell^{XY,z_i} = \frac{\left|C_\ell^{XY,z_i} - \tilde{C}_\ell^{XY}\right|}{\sigma_\ell^{XY,z_i}},
\end{equation}
and $\tilde{C}_\ell^{XY}$ is the unlensed power spectrum.

We use the above to calculate an approximate lensing SNR per redshift and {\it{comoving wavenumber~$k$}}.  The wavenumber~$k$ corresponding to the multipole~$\ell$ at the redshift~$z_i$ is given by $k \approx (\ell + 1/2) / \chi(z_i)$.  To make the bottom panel of Fig.~\ref{fig:SNR}, we first bin the lensed spectra $C_\ell^{XY,z_i}$, unlensed spectra $\tilde{C}_\ell^{XY}$, and diagonal covariance matrices $\mathbb{C}^{XY,XY;z_i}_{\ell \ell'}$ using uniform $\Delta \ell = 30$ binning. We denote the upper and lower multipoles of the $b^\mathrm{th}$ multipole bin by $\ell_{b,\mathrm{min}}$ and $\ell_{b,\mathrm{max}}$, respectively, and the bin center by $\ell_b$.  This results in a SNR per redshift bin~$i$ and per multipole bin~$b$, $(S/N)^{z_i}_{\ell_b}$. We then assign this lensing SNR value to the corresponding wavenumber range, from $k_\mathrm{min}(\ell_b, z_i) = (\ell_{b,\mathrm{min}} + 1/2) / \chi(z_i)$ to $k_\mathrm{max}(\ell_b, z_i) = (\ell_{b,\mathrm{max}} + 1/2) / \chi(z_i)$.

In the top panel of Fig.~\ref{fig:SNR}, we plot the lensing SNR per wavenumber bin, summed over the redshift bins in the range $0 \leq z \leq 6$. We consider 11 bins in wavenumber $k$, evenly spaced in $\log_{10}(k)$, denoting the upper and lower edges of the $j^\mathrm{th}$ bin by $k_{j,\mathrm{min}}$ and $k_{j,\mathrm{max}}$, respectively, and the bin center by $k_j$.  For the $i^\mathrm{th}$ redshift bin and $j^\mathrm{th}$ wavenumber bin, we apply the approximation $k \approx (\ell + 1/2) / \chi(z)$ to calculate the corresponding multipole range, $\ell_\mathrm{min}(z_i,k_j) = \chi(z_i) k_{j,\mathrm{min}} - 1/2$ to $\ell_\mathrm{max}(z_i, k_j) = \chi(z_i) k_{j,\mathrm{max}} - 1/2$. The lensing SNR in the $i^\mathrm{th}$ redshift bin and $j^\mathrm{th}$ wavenumber bin is given by summing $(S/N)_\ell^{z_i}$ in quadrature over this multipole range 
\begin{equation}
        \left(\frac{S}{N}\right)^{z_i}_{k_j} = \sqrt{\sum_{\ell=\ell_\mathrm{min}(z_i,k_j)}^{\ell_\mathrm{max}(z_i,k_j)} \left[\left(\frac{S}{N}\right)^{z_i}_\ell\right]^2}.
\end{equation}
We then sum over the redshift bins 
\begin{equation} 
        \left(\frac{S}{N}\right)_{k_j} = \sqrt{\sum_i \left[ \left(\frac{S}{N}\right)^{z_i}_{k_j}\right]^2}
\end{equation}
to obtain a lensing SNR for each wavenumber bin~$j$.

For each wavenumber bin shown in the top panel of Fig.~\ref{fig:SNR}, we show the contribution from each redshift bin to the total lensing SNR in that wavenumber bin. We quantify this by taking the ratio $[(S/N)^{z_i}_{k_j}]^2 / [(S/N)_{k_j}]^2$, which gives the fractional contribution of the $i^\mathrm{th}$ redshift bin to the total lensing SNR in the $j^\mathrm{th}$ wavenumber bin, and then multiplying this by the total lensing SNR $(S/N)_{k_j}$ in that wavenumber bin.

To obtain the error bars in Fig.~\ref{fig:Pk}, we follow the same procedure to obtain a SNR value in each wavenumber bin~$j$, $(S/N)_{k_j}$, but instead use 22 wavenumber bins. We place the data points on the theoretical linear matter spectrum at the $k$-bin centers, and use 
\begin{equation} 
    \begin{split}
            \sigma(P_\mathrm{m}^\mathrm{lin}(k_j)) & = \frac{P_\mathrm{m}^\mathrm{lin}(k_j)}{(S/N)_{k_j}}
    \end{split}
\end{equation} 
to calculate the CMB-HD error bars.

\begin{table}[t]
    \begin{center}
    \begin{tabular}{l @{\hskip 1em}c @{\hskip 1em}c  @{\hskip 2em}c @{\hskip 1em}c}
      \toprule
      \toprule
       & \multicolumn{4}{c}{Lensing SNR}
      \\
      \cmidrule(){2-5}
        & \multicolumn{2}{c@{\hskip 2em}}{$z \in [0, 1100]$} &  \multicolumn{2}{c}{$z \in [0, 6]$}
      \\
      Spectra  & correct  & approx.  & correct & approx.
      \\
      \midrule
      $TT$-only  & 1573 & 850 & 1353 & 770
      \\
      $TE$-only  & 161 & 103 & 133 & 90
      \\
      $EE$-only  & 276 & 149 & 221 & 126
      \\
      $BB$-only  & 493 & 1044 & 386 & 906
      \\
      $\kappa\kappa$-only  & 1287  & 1178 & 1029 & 1037
      \\
      All & \textbf{1947} & 1798 & 1632 & 1585
      \\
      \bottomrule
    \end{tabular}
    \caption{We compare the CMB-HD lensing SNR obtained from Eq.~\ref{eq:SNR} (columns labelled "correct") and from Eq.~\ref{eq:SNRsumOverEllandzSimple} (equivalent to Eq.~\ref{eq:SNperzAndell} after summing over the redshifts and multipoles), where we apply the approximations described in Appendix~\ref{sec:SNRapprox} (columns labelled "approx.").  We use either the full redshift range $z \in [0,1100]$ back to recombination, or the range $z \in [0,6]$ (used in Figures~\ref{fig:Pk} and~\ref{fig:SNR}).  We list the lensing SNRs for each individual spectrum $XY \in \{TT,TE,EE,BB,\kappa\kappa\}$ in the first five rows, and for the combination of all the spectra in the last row, calculated over the multipole ranges listed in Table~\ref{tab:ExpConfig}. The SNRs obtained from Eq.~\ref{eq:SNR} use the full covariance matrix (or the $XY \times XY$ block for individual spectra), including off-diagonal elements, with spectra obtained by integrating Eq.~\ref{eq:ClPhiPhi} over the given redshift range, as described in Section~\ref{sec:signal}. When using Eq.~\ref{eq:SNRsumOverEllandzSimple}, we calculate the spectra and a corresponding diagonal covariance matrix within each redshift bin, and then take a sum over redshifts.  Note that the first column contains the same information as Table~\ref{tab:lensing_snr}. We emphasize that the approximate lensing SNR calculations are only used in Figures~\ref{fig:Pk} and~\ref{fig:SNR} for illustrative purposes. }
    \label{tab:lensing_snr_approx}
    \end{center}
\end{table}

{\bf{Approximations approximately cancelling:}} In general, neglecting the off-diagonal components of a covariance matrix will result in overestimating the total SNR. However, summing the SNR from redshift bins in quadrature also underestimates the SNR.  We see the latter from 
\begin{equation} \label{eq:SNRsumOverEllSimple}
    \begin{split}
        \frac{S}{N} & = \sqrt{\sum_{\ell} \left[\left(\frac{S}{N}\right)_\ell\right]^2 } \\ & = \sqrt{\sum_{\ell} \left(\frac{C_\ell - \tilde{C}_\ell}{\sigma_\ell}\right)^2 } \\ & = \sqrt{\sum_{\ell} \left(\frac{\sum_i \Delta C_\ell^{z_i}}{\tilde{\sigma}_\ell + \sum_i \Delta \sigma_\ell^{z_i}}\right)^2 }.
    \end{split}
\end{equation}
where we have assumed a diagonal covariance matrix with
\begin{equation}
    \begin{split}
        \tilde{\sigma}_\ell + \Delta \sigma_\ell^{z_i} & = \sqrt{\frac{2}{f_\mathrm{sky} (2\ell + 1)}} \left(\tilde{C}_\ell + \Delta C_\ell^{z_i} + N_\ell\right) .
    \end{split}
\end{equation}
Here $i$ sums over redshift, and $\tilde{C}_\ell$ is the unlensed CMB.
If instead we add in quadrature the SNR from each  redshift bin we have
\begin{equation} \label{eq:SNRsumOverEllandzSimple}
    \begin{split}
         \sqrt{\sum_i \sum_{\ell} \left[\left(\frac{S}{N}\right)^{z_i}_\ell\right]^2 } & =  \sqrt{\sum_i \sum_{\ell} \left(\frac{C_\ell^{z_i} - \tilde{C}_\ell}{\tilde{\sigma}_\ell + \Delta \sigma_\ell^{z_i}}\right)^2 } \\  & =  \sqrt{\sum_{\ell} \sum_i  \left(\frac{\Delta C_\ell^{z_i}}{\tilde{\sigma}_\ell }\right)^2 } 
    \end{split}
\end{equation}
in the limit where $\tilde{\sigma}_\ell \gg \sum_i \Delta \sigma_\ell^{z_i}$, or equivalently $\tilde{C}_\ell~+~N_\ell \gg \sum_i \Delta C_\ell^{z_i}$. In this limit, Eq.~\ref{eq:SNRsumOverEllandzSimple} is less than Eq.~\ref{eq:SNRsumOverEllSimple}. However,  we find that the two approximations mentioned above roughly cancel, as we show in Table~\ref{tab:lensing_snr_approx}, and that the agreement between calculations is close.  We show this in Table~\ref{tab:lensing_snr_approx} for both the case where we sum over all redshifts $z \in [0,1100]$ and for $z \in [0,6]$ (used in Figures~\ref{fig:Pk} and~\ref{fig:SNR}).

\bibliographystyle{apsrev4-1}
\bibliography{main.bib}

\end{document}